\documentclass[twocolumn,preprintnumbers]{revtex4}

\usepackage{graphicx}
\usepackage{epsf}
\usepackage{amsmath,amssymb}
\newcommand{\bvec}{\boldsymbol}
\newcommand{\La}{\Lambda}
\newcommand{\LLa}{{\Lambda\Lambda}}
\newcommand{\LZ}{^{A-1}_{\ \ \ \La} Z}
\newcommand{\LLZ}{^{\ \,A}_{\La\La} Z}
\newcommand{\LLHe}{^{\ \ 6}_{\La\La} \textrm{He}}
\newcommand{\LLBe}{^{\ 10}_{\La\La} \textrm{Be}}

\begin{document}
%\preprint{KUNS-2503}
\title{Structures of $p$-shell double-$\Lambda$ hypernuclei
studied with microscopic cluster models}

\author{Yoshiko Kanada-En'yo}
\affiliation{Department of Physics, Kyoto University, Kyoto 606-8502, Japan}
\begin{abstract}
$0s$-orbit $\Lambda$ states 
in $p$-shell  double-$\Lambda$ hypernuclei ($\LLZ$), 
$^{\ \,8}_{\Lambda\Lambda}\textrm{Li}$, $^{\ \,9}_{\Lambda\Lambda}\textrm{Li}$, $^{10,11,12}_{\ \  \ \ \ \Lambda\Lambda}\textrm{Be}$, 
$^{12,13}_{\ \ \Lambda\Lambda}\textrm{B}$, and $^{\,14}_{\Lambda\Lambda}\textrm{C}$ are investigated.
Microscopic cluster models are applied to core nuclear part and a potential
model is adopted for $\Lambda$ particles. The $\Lambda$-core potential is a folding 
potential  obtained with effective $G$-matrix $\Lambda$-$N$  interactions, which 
reasonably reproduce energy spectra of $\LZ$. System dependence of the
$\La$-$\La$ binding energies is  understood by the core polarization energy from nuclear size reduction. 
Reductions of nuclear sizes and $E2$ transition strengths by $\Lambda$ particles are also discussed. 
%The results reproduce well the experimental
%$E2$ transition strengths in $^6\textrm{Li}$ and $^7_\Lambda \textrm{Li}$ without using effective charges. 
\end{abstract}
\maketitle

\section{Introduction}
In the recent progress in strangeness physics, experimental and theoretical 
studies on hypernuclei have been extensively  performed. 
Owing to experiments with high-resolution $\gamma$-ray measurements, 
detailed spectra of $p$-shell $\Lambda$ hypernuclei have been experimentally revealed
\cite{Hashimoto:2006aw,Tamura:2010zz,Tamura:2013lwa}. 
The data of energy spectra and electromagnectic transitions have been providing useful information for properties of 
$\Lambda$-nucleon($N$) interactions. They are also helpful to investigate impurity effects of a $\Lambda$ particle 
on nuclear systems. 
In a theoretical side, 
structure studies of $p$-shell $\Lambda$ hypernuclei have been performed with various models such as 
cluster models \cite{Motoba:1984ri,motoba85,Yamada:1985qr,Yu:1986ip,Hiyama:1996gv,Hiyama:1997ub,Hiyama:1999me,Hiyama:2000jd,Hiyama:2002yj,Hiyama:2006xv,Hiyama:2010zzc,Cravo:2002jv,Suslov:2004ed,Mohammad:2009zza,Zhang:2012zzg,Funaki:2014fba,Funaki:2017asz}, shell models \cite{Gal:1971gb,Gal:1972gd,Gal:1978jt,Millener:2008zz,Millener:2010zz,Millener:2012zz}, mean-field and beyond mean-field models \cite{Guleria:2011kk,Vidana:2001rm,Zhou:2007zze,Win:2008vw,Win:2010tq,Lu:2011wy,Mei:2014hya,Mei:2015pca,Mei:2016lce,Schulze:2014oia}, hyper antisymmetrized molecular dynamics (HAMD) model
\cite{Isaka:2011kz,Isaka:2015xda,Homma:2015kia,Isaka:2016apm,Isaka:2017nuc},  and no-core shell model \cite{Wirth:2014apa}, and so on.

For double-$\La$ hypernuclei, experimental observations 
with  nuclear emulsion have been used
to extract information of the $\La$-$\La$ interaction in nuclear systems 
from binding energies \cite{Nakazawa:2010zzb}. 
Several double-$\La$ hypernuclei have been observed, but precise data of binding energies
are still limited because of experimental uncertainties in energies and 
reaction assignments. 
The most reliable datum is the binding energy of $\LLHe$, which suggests 
an weak attractive $\Lambda$-$\Lambda$ interaction. Another observation is a candidate event for
$\LLBe^*$. 
In order to extract information of the $\La$-$\La$ interaction from the limited experimental data, systematic investigations of 
binding energies of $p$-shell $\LZ$ and $\LLZ$ have been performed 
with semi-microscopic cluster model  \cite{Hiyama:2002yj,Hiyama:2010zzd,Hiyama:2010zz} and shell model
calculations \cite{Gal:2011zr}. 
In the former calculation, dynamical effects in three-body and four-body cluster systems 
as well as spin-dependent contributions
are taken into account in a semi-microscopic treatment of antisymmetrization effect between 
clusters called the orthogonal condition model (OCM).
In the latter calculation, 
the spin-dependent and $\Lambda$-$\Sigma$
coupling contributions in $\LZ$ are taken into account perturbatively. 
A major interest concerning the $\La$-$\La$ interaction is so-called 
 $\La$-$\La$ binding energy $\Delta B_\LLa$ defined with masses $M$ of $^{A-2}Z$, $\LZ$, and $\LLZ$ as, 
\begin{equation}
\Delta B_\LLa(\LLZ)=2M(\LZ) - M(^{A-2}Z)-M(\LLZ), 
\end{equation}
which stands for the difference of the two-$\Lambda$ binding energy in $\LLZ$ 
from twice of the single-$\Lambda$ binding energy in $\LZ$. In Refs.~\cite{Hiyama:2002yj,Hiyama:2010zzd,Gal:2011zr}, 
they have discussed systematics of 
$\Delta B_\LLa(\LLZ)$ comparing with available data, and pointed out that 
$\Delta B_\LLa(\LLZ)$ has rather strong system (mass-number) dependence because of various effects and
is not a direct measure of the $\La$-$\La$ interaction. For example, 
$\Delta B_\LLa(\LLZ)$ of $\LLBe$ is significantly deviated from 
global systematics in $p$-shell double-$\La$ hypernuclei because of remarkable
$2\alpha$ clustering in the core nucleus, $^8\textrm{Be}$. 

In our previous work \cite{Kanada-Enyo:2017ynk}, we have investigated
energy spectra of low-lying $0s$-orbit $\Lambda$ states 
in $p$-shell  $\Lambda$ hypernuclei  by applying 
microscopic cluster models for core nuclei and a single-channel
potential model for a $\Lambda$ particle. As the core polarization effect in $\LZ$,
the nuclear size reduction contribution by a $\Lambda$ has been taken into account. 
The $\Lambda$-core potentials have been calculated with local density approximations
of folding potentials using the $G$-matrix $\Lambda$-$N$ interactions. 
Since the spin-dependence of the $\Lambda$-$N$ interactions are ignored, 
the spin-averaged energies of low-energy spectra have been discussed. 
The previous calculation describes systematic trend of experimental data
for excitation energy shift by a $\Lambda$ from $^{A-1} Z$ to $^A_\Lambda Z$, and 
shows that nuclear size difference between the ground and excited states dominantly 
contributes to the excitation energy shift. 
The size reduction by a $\Lambda$ particle in $\LZ$ has been also studied, and 
it was found that significant size reduction occurs
in $^7_\Lambda \textrm{Li}$ and $^9_\Lambda \textrm{Be}$ because of 
developed clustering consistently with 
predictions by other calculations \cite{Motoba:1984ri,Hiyama:1999me,Funaki:2014fba}. 
The framework developed in the previous work of $\La$ hypernuclei 
can be applied also to double-$\La$ hypernuclei
straightforwardly.

In the present paper, we extend the previous calculation to $p$-shell double-$\La$ hyper nuclei.
We solve motion of two $S$-wave $\Lambda$ particles around a core nucleus
in the $\La$-core potential calculated by folding the
 $G$-matrix $\Lambda$-$N$ interactions in the same way as 
the previous calculation for $\LZ$. For the effective $\Lambda$-$\Lambda$ interaction, we 
adopt the parametrization used in Refs.~\cite{Hiyama:2002yj,Hiyama:2010zzd}
with a slightly modification to reproduce the $\Lambda$-$\Lambda$ binding energy in $\LLHe$. 
We investigate systematics of the $\La$-$\La$ binding energies in $p$-shell double-$\Lambda$ hypernuclei.  
System dependence of the $\La$-$\La$ binding energies is discussed in relation with the core polarization. 
We also discuss the size reduction and excitation energy shift by $\Lambda$ particles in $\LZ$ and $\LLZ$.  

In the present calculation, $S$-wave $\Lambda$s are assumed and the spin dependence of the $\La$-$N$ and $\La$-$\La$ 
interactions are disregarded. 
In $\LLZ$, the total spin $J^\pi$ of $\LLZ$ is given just by the core nuclear spin
$I^\pi$ as $\LLZ(J^\pi=I^\pi)$.  
In $\LZ$ with a non-zero nuclear spin ($I\ne 0$), 
spin-doublet $J^\pi= (I\pm 1/2)^\pi$ states completely degenerate in the present calculation. 
For simplicity, we denote the spin-doublet $J^\pi= (I\pm 1/2)^\pi$ states  
in $\Lambda$ hypernuclei with the label  $I^\pi$ of the core nuclear spin  as 
$\LZ(I^\pi)$. 

This paper is organized as follows. In the next section, we explain the framework of the present calculation.
The effective $N$-$N$, $\Lambda$-$N$, and  $\Lambda$-$\Lambda$ 
interactions are explained in Sec.~\ref{sec:interactions}.
Results and discussions are given in Sec.~\ref{sec:results}.
Finally, the paper is summarized in Sec.~\ref{sec:summary}.

\section{Framework}\label{sec:formulation}
\subsection{Microscopic cluster models for core nuclei}
Core nuclei $^{A-2}Z$ in $\LZ$ and $\LLZ$
are calculated with the microscopic cluster models in the same way as the previous
calculation for $\LZ$ \cite{Kanada-Enyo:2017ynk}.
The generator coordinate method (GCM) \cite{Hill:1952jb,Griffin:1957zza} is applied using the Brink-Bloch cluster wave functions \cite{Brink66}
of $\alpha+d$, $\alpha+t$, $2\alpha$, $2\alpha+n$, $2\alpha+nn$, $2\alpha+d$, $2\alpha+t$, and $3\alpha$
clusters for $^6\textrm{Li}$,  $^7\textrm{Li}$, $^8\textrm{Be}$,  $^{9}\textrm{Be}$,   $^{10}\textrm{Be}$, $^{10}\textrm{B}$, 
 $^{11}\textrm{B}$, and  $^{12}\textrm{C}$, respectively.
$d$, $nn$, $t$, and $\alpha$ clusters are written by harmonic oscillator 
$0s$ configurations with a common width parameter $\nu=0.235$ fm$^{-2}$.  
For  $^{11}\textrm{B}$ and  $^{12}\textrm{C}$, the $p_{3/2}$ configurations are
taken into account by adding the corresponding shell model wave functions 
to the $2\alpha+t$, and $3\alpha$ cluster wave functions as done in Refs.~\cite{Kanada-Enyo:2017ynk,Suhara:2014wua}. 
For $^{10}$Be,  the $^6\textrm{He}+\alpha$ wave functions adopted in Ref.~\cite{Kanada-Enyo:2016jnq} are
superposed in addition to the $2\alpha+nn$ wave functions. 

Let us consider a $A_N$-nucleon system for a mass number $A_N=A-2$ nucleus consisting of $C_1,\ldots,C_k$ clusters. 
$k$ is the number of clusters. 
The Brink-Bloch cluster wave function $\Phi_\textrm{BB}(\bvec{S}_1,\ldots, \bvec{S}_k)$
is written by a $A_N$-body microscopic wave function parametrized by the 
cluster center parameters $\bvec{S}_j$ ($j=1,\ldots,k$). 
To take into account inter-cluster motion,  the GCM is applied to the spin-parity projected  
Brink-Bloch cluster wave functions with respect to the generator coordinates $\bvec{S}_j$.
Namely, the wave function $\Psi(J^\pi_n)$  for the 
$J^\pi_n$ state is given by a linear combination of the 
Brink-Bloch wave functions with various configurations of $\{\bvec{S}_1,\ldots, \bvec{S}_k\}$ as 
\begin{equation}\label{eq:gcm-wf}
\Psi(J^\pi_n)=\sum_{\bvec{S}_1,\ldots, \bvec{S}_k} \sum_{K} c^{J^\pi_n}_{\bvec{S}_1,\ldots, \bvec{S}_k,K}
P^{J\pi}_{MK} \Phi_\textrm{BB}(\bvec{S}_1,\ldots, \bvec{S}_k),
\end{equation}
where $P^{J\pi}_{MK}$ is the spin-parity projection operator. The coefficients 
$c^{J^\pi_n}_{\bvec{S}_1,\ldots, \bvec{S}_k,K}$ are determined by solving Griffin-Hill-Wheeler equations 
\cite{Hill:1952jb,Griffin:1957zza}, which is equivalent to the diagonalization of the Hamiltonian and norm matrices.
For the $\alpha+d$ and $2\alpha$ wave functions, $\bvec{S}_1$ and  $\bvec{S}_2$  are chosen to be 
$\bvec{S}_1-\bvec{S}_2=(0,0,d)$ with $d=\{1,2,\cdots,15$ fm\}. 
For the $\alpha+t$ wave functions,  $d=\{1,2,\cdots,8$ fm\} are adopted to obtain a bound state solution for 
the resonance state $^7\textrm{Li}(7/2^-_1)$ corresponding to a bound state approximation. 
%The effect of this truncation 
%gives only minor effect to the bound states, $^7\textrm{Li}(3/2^-_1)$ and $^7\textrm{Li}(1/2^-_1)$. 
For the configurations of $2\alpha+nn$, $2\alpha+d$, $2\alpha+t$, and $3\alpha$, $\bvec{S}_{1,2,3}$ are chosen to be 
\begin{eqnarray}
&&\bvec{S}_1-\bvec{S}_2=(0,0,d), \\
&&\bvec{S}_3-\frac{A_2\bvec{S}_1+A_1\bvec{S}_2}{A_1+A_2}
=(r\sin \theta,0,r \cos \theta),
\end{eqnarray}
with $d=\{1.2,2.2,\ldots,4.2$ fm\}, $r=\{0.5,1.5,\ldots,4.5$ fm\}, and $\theta=\{0, \pi/8, \ldots, \pi/2\}$. 
Here $A_i$ is the mass number of the $C_i$ cluster.
For the $2\alpha+n$  cluster, a larger model space of 
$d=\{1.2,2.2,\ldots,6.2$ fm\}, $r=\{0.5,1.5,\ldots,6.5$ fm\}, and 
$\theta=\{0, \pi/8, \ldots, \pi/2\}$ are used to describe remarkable clustering  in $^9$Be. 

The Hamiltonian of the nuclear part consists of the kinetic term, effective nuclear interactions, and Coulomb interactions
as follows, 
\begin{eqnarray}
H_N&=&T+V^\textrm{(c)}_N+V^\textrm{(so)}_N+V_\textrm{coul},\\
T&=&\sum^{A_N}_{i} \frac{1}{2m_N}\bvec{p}^2_i -T_G,\\
V^\textrm{(c)}_N&=&\sum^{A_N}_{i<j} v^{(c)}_{NN}(i,j),\\
V^\textrm{(so)}_N&=&\sum^{A_N}_{i<j} v^{\textrm{(so)}}_{NN}(i,j),\\ 
V_\textrm{coul}&=&\sum^Z_{i<j} v_\textrm{coul}(r_{ij}),
\end{eqnarray}
where  the kinetic term of the center of mass (cm) motion, $T_G$, is subtracted exactly.  
$v^{(c)}_{NN}(i,j)$ and  $v^{\textrm{(so)}}_{NN}(i,j)$ are the effective $N$-$N$ central and spin-orbit  interactions, respectively.
The nuclear energy $E_N=\langle \Psi(J^\pi_n) |H_N|\Psi(J^\pi_n) \rangle$ 
and nuclear density  $\rho_N(r)$ of the core nuclei are calculated 
for the obtained GCM wave function $\Psi(J^\pi_n)$.
Here the radial coordinate $r$ in $\rho_N(r)$ is the distance measured from the cm of core nuclei. 

\subsection{Folding potential model for $\Lambda$ particles}
Assuming two $0s$-orbit $\La$ particles coupling to the spin-singlet state, 
$(0s)^2_\Lambda$ states of  $\LLZ$ are calculated with a folding potential model
in a similar way to the previous calculation for $\LZ$ \cite{Kanada-Enyo:2017ynk}. 

The Hamiltonian of $(0s)_\Lambda$ states in $\LZ$ is given by 
the nuclear Hamiltonian $H_N$ for the core nucleus $^{A-2}Z$ and the single-particle Hamiltonian 
$h_{\Lambda C}$ 
for a $\Lambda$ particle around the core as 
\begin{eqnarray}
H&=&H_N+h_{\Lambda C},\\
%\end{eqnarray}
%where the single-particle Hamiltonian  $h_{\Lambda C}$ 
%for a $\Lambda$ particle around the core nuclei $^{A-2}_\La Z$
%in $^{A-1}_\La Z$ is  
%\begin{eqnarray}
h_{\Lambda C}&=&t_{\Lambda}+U_{\Lambda C}, \\
t_{\Lambda}&=&\frac{1}{2\mu_\Lambda}\bvec{p}^2, \\
\mu_\Lambda&=&\frac{(A-2)m_N m_\Lambda}{(A-2)m_N + m_\Lambda}.
\end{eqnarray}
%$\bvec{r}$, $\bvec{r}'$, and $\bvec{p}$ are defined with respect to the relative coordinate of the $\Lambda$ 
%from the cm of the core nucleus. 

The Hamiltonian of $(0s)^2_\Lambda$ states in $\LLZ$ is written straightforwardly as 
\begin{eqnarray}
H&=&H_N+H_{\Lambda\Lambda}\\
H_{\Lambda\Lambda}&=&h_{\Lambda C}(1)+h_{\Lambda C}(2)+V_{\La\La}(r_{12}),
\end{eqnarray}
where $V_{\La\La}$ is the even-part of the $\Lambda$-$\Lambda$ interactions. Here, 
the recoil kinetic term is dropped off for $(0s)^2_\Lambda$ states. 

The $\Lambda$-core potential $U_{\Lambda C}$ is calculated by folding the $\Lambda$-$N$ interactions with 
the nuclear density $\rho_N(r)$ as 
\begin{eqnarray}
U_{\La C}(\bvec{r},\bvec{r}')&=&U^\textrm{D}_{\La C}(\bvec{r})
+ | \bvec{r} \rangle  U^\textrm{EX}_{\La C}(\bvec{r},\bvec{r}')\langle \bvec{r}'|, \\
U^\textrm{D}_{\La C}(\bvec{r})&=&\int \bvec{r}'' \rho_N(\bvec{r}'') v^\textrm{D}_{\Lambda N}(k_f; |\bvec{r}-\bvec{r}''|),\\
U^\textrm{EX}_{\La C}(\bvec{r},\bvec{r}')&=&
\rho_N(\bvec{r},\bvec{r}') v^\textrm{EX}_{\Lambda N}(k_f; |\bvec{r}-\bvec{r}'|),\\
 v^\textrm{D}_{\Lambda N}(k_f; r)&=&\frac{1}{2}\left[ V^\textrm{e}_{\Lambda N}(k_f; r)+ V^\textrm{o}_{\Lambda N}(k_f; r)\right],\\
 v^\textrm{EX}_{\Lambda N}(k_f; r)&=&\frac{1}{2}\left[ V^\textrm{e}_{\Lambda N}(k_f; r)- V^\textrm{o}_{\Lambda N}(k_f; r)\right],
\end{eqnarray}
where
$ V^\textrm{e}_{\Lambda N}(k_f; r)$  and  $V^\textrm{o}_{\Lambda N}(k_f; r)$ are even and odd parts of the
effective $\Lambda$-$N$ central interactions, respectively, including the parameter $k_f$ for density dependence.
The nuclear density matrix $\rho_N(\bvec{r},\bvec{r}')$ 
in the exchange potential  $U^\textrm{EX}_{\La C}(,\bvec{r}')$ is approximated
with the density matrix expansion using the local density approximation  \cite{Negele:1975zz} as done in
the previous paper. %Ref.~\cite{Kanada-Enyo:2017ynk}.

For a given nuclear density $\rho_N(r)$ of a core nuclear state of $^{A-2} Z$, 
the $\Lambda$-core wave functions $\phi_\Lambda(r)$ in 
 $\LZ$ and  $\LLZ$ are 
calculated with the Gaussian expansion method \cite{Kamimura:1988zz,Hiyama:2003cu} so as to minimize the single-$\La$ 
energy $E_{\Lambda}= \langle\phi_\Lambda| h_{\La C} |\phi_\Lambda\rangle$
and two-$\Lambda$ energy $E_{\Lambda\Lambda}= \langle\phi^2_\Lambda| H_{\La\La} |\phi^2_\Lambda\rangle$, respectively.
The rms radius ($r_\Lambda$) and the averaged nuclear density ($\langle {\rho}_N\rangle_\Lambda$)
for the $\Lambda$ distribution are calculated with the obtained $\Lambda$-core wave function $\phi_\Lambda(r)$ , 
\begin{eqnarray}
r_\Lambda&=&\sqrt{\int  \phi_\Lambda^*(r)\phi_\Lambda(r) r^2 d\bvec{r}},\\
\langle {\rho}_N\rangle_\Lambda&=&\int  \phi_\Lambda^*(r)\phi_\Lambda(r) \rho_N(r)  d\bvec{r}.
\end{eqnarray}

\subsection{Core polarization}
We take into account the core polarization, {\it i.e.}, the nuclear structure change induced by $0s$-orbit $\Lambda$s, 
in the same way as done in the previous calculation of $\Lambda$ hypernuclei. 
To prepare  nuclear wave functions polarized by the $\Lambda$ particles,
we add artificial nuclear interactions $\epsilon V^\textrm{art}$ to the nuclear Hamiltonian. By 
performing the GCM cluster-model calculation of the core nuclear part
for the modified Hamiltonian $H_N+\epsilon V^\textrm{art}$, 
we obtain wave function $\Phi_N(\epsilon)$ for $^{A-2}Z$. For the prepared nuclear wave function,  we calculate 
the nuclear energy $E_N(\epsilon)=\langle \Phi_N(\epsilon)|H_N |\Phi_N(\epsilon)\rangle$ for the original 
nuclear Hamiltonian $H_N$ of $^{A-2}Z$ without the artificial nuclear interactions.
Using the nuclear density $\rho_N(\epsilon;r)$ obtained with $\Phi_N(\epsilon)$, 
the single- and two-$\La$ energies ($E_\La(\epsilon)$ and $E_\LLa(\epsilon)$) in $\LZ$ and  $\LLZ$ are
calculated. Finally, the optimum $\epsilon$  value, {\it i.e.}, the optimum nuclear wave function $\Phi_N(\epsilon)$ is chosen 
for each $I^\pi$ state in $\LZ$ and $\LLZ$ 
so as to minimize the total energy
\begin{eqnarray}
E(\epsilon;\LZ)&=&E_N(\epsilon)+E_\Lambda(\epsilon)
\end{eqnarray}
in $\LZ$, and 
\begin{eqnarray}
E(\epsilon;\LLZ)&=&E_N(\epsilon)+E_{\La\La}(\epsilon)
\end{eqnarray}
in $\LLZ$.

For the artificial interaction $V^\textrm{art}$, we use the central part of the nuclear interactions as
$V^\textrm{art}=V^\textrm{(c)}_N$. It corresponds to slight enhancement of the central nuclear interaction
as 
\begin{equation}
H_N+\epsilon V^\textrm{art} =T+(1+\epsilon)V^\textrm{(c)}_N+V^\textrm{(so)}_N+V_\textrm{coul},\\
\end{equation}
where $\epsilon(\ge 0)$ is regarded as the enhancement factor to simulate the nuclear structure change 
induced by the $0s$-orbit $\Lambda$ particles.
The main effect of the enhancement 
on the structure change is size reduction of core nuclei. Therefore, 
it is considered, in a sense, that 
the present treatment of the core polarization simulates 
the nuclear size reduction, 
which is determined by energy balance between the $\Lambda$ potential energy gain and 
nuclear energy loss.  
In the optimization of $\epsilon$, we vary only the GCM coefficients 
but fix the basis cluster wave functions,
corresponding to the inert
cluster ansatz. In this assumption, the enhancement of the central nuclear interactions acts as an enhancement of the 
inter-cluster potentials. 

\subsection{Definitions of energies and sizes for $\LZ$ and $\LLZ$ systems}
The $\Lambda$ binding energy ($B_\Lambda$) in  $\LZ$
is calculated as 
\begin{eqnarray}
B_\Lambda(\LZ)&=&-\left[E(\epsilon;\LZ)-E_N(\epsilon=0)\right]\nonumber\\
&=&-\left[\delta_\Lambda(E_N)+E_\Lambda \right], \label{eq:BL} \\ 
\delta_\Lambda(E_N)&\equiv& E_N(\epsilon;\LZ)-E_N(\epsilon=0),
\end{eqnarray}
where  $E_N(\epsilon=0)$ is the unperturbative nuclear energy without the $\Lambda$ particle
and $\delta_\Lambda(E_N)$ stands for the nuclear energy increase by a  $\Lambda$ particle 
in $\LZ$. 

Similarly, the two-$\La$ binding energy ($B_\LLa$) in  $\LLZ$
is calculated as 
\begin{eqnarray}
B_\LLa(\LLZ)&=&-\left[E(\epsilon;\LLZ)-E_N(\epsilon=0)\right]\nonumber\\
&=&-\left[\delta_\LLa(E_N)+E_\LLa \right], \\ 
\delta_\LLa(E_N)&\equiv& E_N(\epsilon;\LLZ)-E_N(\epsilon=0),
\end{eqnarray}
where $\delta_\LLa(E_N)$ is the nuclear energy increase caused by two $\Lambda$ particles 
in $\LLZ$. 

The $\La$-$\La$ binding energy $\Delta B_\LLa$  in $\LLZ$ is given as 
\begin{eqnarray}
&&\Delta B_\LLa(\LLZ)=B_\LLa(\LLZ)-2B_\La(\LZ)\\
&&=-\left[ E(\LLZ)+E(^{A-2}Z)-2E(\LZ)\right]. 
\end{eqnarray}
To discuss the $\LLa$ binding, we also define $\Delta B_\LLa$  for excited states ($I^\pi_n$) as
\begin{eqnarray}
&&\Delta B_\LLa(\LLZ;I^\pi_n)\nonumber\\
&&=-\left[ E(\LLZ;I^\pi_n)+E(^{A-2}Z;I^\pi_n)-2E(\LZ;I^\pi_n)\right]. \nonumber\\
\end{eqnarray}
As discussed by Hiyama {\it et al.} in Refs.~\cite{Hiyama:2002yj,Hiyama:2010zzd}, 
$\Delta B_\LLa$ is not necessarily a direct measure of the $\Lambda$-$\Lambda$ interaction because it 
is contributed also by various effects such as the core polarization effect. Alternatively, 
the $\LLa$ bond energy ${\cal V}^\textrm{bond}_\LLa$
\begin{eqnarray}\label{eq:vbond}
{\cal V}^\textrm{bond}_\LLa\equiv B_\LLa(\LLZ)-B_\LLa(\LLZ;V_\LLa=0)
\end{eqnarray}
has been discussed in Refs.~\cite{Hiyama:2002yj,Hiyama:2010zzd}. 
Here $B_\LLa(\LLZ;V_\LLa=0)$ is the two-$\La$ binding energy 
obtained by switching off the $\La$-$\La$ interactions ($V_\LLa=0$), and 
${\cal V}^\textrm{bond}_\LLa$
indicates the difference of $B_\LLa$ between calculations
with and without the $\La$-$\La$ interactions.

For excited states $\LZ(I^\pi_n)$ and $\LLZ(I^\pi_n)$ 
with the excited core $^{A-2}Z(I^\pi_n)$, 
excitation energies measured from the ground states 
are denoted by  $E_x(\LZ;I^\pi_n)$ 
and $E_x(\LLZ;I^\pi_n)$, respectively. 
Note that the parameter $\epsilon$ is optimized for each state
meaning that the core polarization is state dependent.
Because of the impurity effect of $\Lambda$ particles, 
the excitation energies are changed from the original
excitation energy $E_x(^{A-2}Z;I^\pi_n)$ of isolate nuclei $^{A-2}Z$.
For each $I^\pi$, we denote the energy change caused by $\Lambda$s as 
\begin{eqnarray}
\delta_\La(E_x)&\equiv&E_x(\LZ)- E_x(^{A-2}Z),\\
\delta_\LLa(E_x)&\equiv&E_x(\LLZ)- E_x(^{A-2}Z),
\end{eqnarray}
which we call the excitation energy shift. 

The nuclear sizes $R_N$
in $^{A-2}Z$, $\LZ$, and $\LLZ$ are calculated with 
the nuclear density $\rho_N(r)$ as 
\begin{eqnarray}
R^2_N&=&\frac{1}{A-2}\int  4\pi r^4 \rho_N(r)  dr.
\end{eqnarray}
We 
define nuclear size changes $\delta_\La (R_N)$ and $\delta_\LLa (R_N)$ 
by $\La$ particles in  $\LZ$ and  $\LLZ$
from the original size as, 
\begin{eqnarray}
\delta_\La (R_N)&=&R_N(\LZ)-R_N(^{A-2}Z),\\
\delta_\LLa (R_N)&=&R_N(\LLZ)-R_N(^{A-2}Z).
\end{eqnarray}

\section{Effective interactions}\label{sec:interactions}
\subsection{Effective nuclear interactions}
We use the same effective two-body  nuclear interactions 
as the previous calculation;
the finite-range central interactions of
the Volkov No.2 force \cite{VOLKOV} with $w=0.40$, $m=0.60$, and $b=h=0.125$
and the spin-orbit interactions of the G3RS parametrization \cite{LS} with $u_1=-u_2=1600$ MeV
except for $^6$Li and $^7$Li.
For $^6$Li, we use the adjusted parameter set, 
$w=0.43$, $m=0.57$, $b=h=0.125$, and $u_1=-u_2=1200$ MeV, which 
reproduces the experimental $^6\textrm{Li}(1^+_1)$ and  $^6\textrm{Li}(3^+_1)$ energies
measured from the $\alpha+d$ threshold energy. The same set of interaction parameters is used also for 
$^7$Li. This interaction set gives the $\alpha+d$ and $\alpha+t$ threshold energies, 1.48 MeV and 2.59 MeV, 
for $^6\textrm{Li}$ and  $^7\textrm{Li}$, respectively.  (The experimental threshold energies are 1.48 MeV for 
$^6\textrm{Li}$ and 2.47 MeV for $^7\textrm{Li}$.)

\subsection{Effective $\Lambda$-$N$ interaction}

As the effective $\Lambda$-$N$ central interactions, 
we use the ESC08a parametrization of the $G$-matrix $\Lambda$-$N$ ($\Lambda NG$) interactions 
derived from $\Lambda$-$N$ interactions of the one-boson-exchange model 
\cite{Yamamoto:2010zzn,Rijken:2010zzb}. Since spin-dependent contributions are ignored in the 
present folding potential model, the $\Lambda$-core potentials are contributed by 
the spin-independent central parts, 
\begin{eqnarray}
V^\textrm{e}_{\Lambda N}(k_f; r)&=&\sum^3_i  (c^\textrm{e}_{0,i}+c^\textrm{e}_{1,i}  k_F+c^\textrm{e}_{2,i}  k_F^2)  \exp\left[-\left(\frac{r}{\beta_i}\right)^2\right],
\nonumber\\
&&\\
V^\textrm{o}_{\Lambda N}(k_f; r)&=&\sum^3_i  (c^\textrm{o}_{0,i}+c^\textrm{o}_{1,i}  k_F+c^\textrm{o}_{2,i}  k_F^2)  \exp\left[-\left(\frac{r}{\beta_i}\right)^2\right],
\nonumber\\
&&\\
c^\textrm{e}_{n,i}&=&\frac{1}{4}c^\textrm{1E}_{n,i}+\frac{3}{4}c^\textrm{3E}_{n,i},\\
c^\textrm{o}_{n,i}&=&\frac{1}{4}c^\textrm{1O}_{n,i}+\frac{3}{4}c^\textrm{3O}_{n,i}. 
\end{eqnarray}
The values of the parameters $\beta_i$ and $c^\textrm{1E,3E,1O,3O}_{n,i}$ are given 
in Table II of Ref.~\cite{Yamamoto:2010zzn}.

As for the $k_f$ parameter in the $\Lambda NG$ interactions,  
we adopt three choices.
The first choice is the density-dependent $k_f$ called ``averaged density approximation (ADA)''  
used in Refs.~\cite{Yamamoto:2010zzn,Isaka:2016apm,Isaka:2017nuc}.
The $k_f$ is taken to be $k_f=\langle k_f \rangle_\Lambda$, where 
$\langle k_f \rangle_\Lambda$ is the averaged Fermi momentum for the $\Lambda$ distribution,
\begin{equation}
\langle k_f \rangle_\Lambda
=\left[\frac {3\pi^2}{2}\langle \rho_N \rangle_\Lambda \right]^{1/3},
\end{equation}
and self-consistently determined for each state.
We label this choice of the $\Lambda NG$ interactions as ESC08a(DD) consistently to the previous paper. 

The second choice is the density-independent interaction with a 
fixed $k_f$ value, $k_f=k^\textrm{inp}_f$. We use a system-dependent but state-independent 
value as the input parameter $k^\textrm{inp}_f$ in calculation of $\LZ$, and use the same value $k^\textrm{inp}_f$ in calculation of $\LLZ$. As the $k^\textrm{inp}_f$ value for each $\LZ$ system, we used 
the averaged value of 
$\langle k_f \rangle_\Lambda$ self-consistently determined by the ADA treatment in the ESC08a(DD) calculation for low-lying states. 
We label this choice as ESC08a(DI). 

The third choice is the hybrid version of the ESC08a(DD) and ESC08a(DI) interactions.
In the previous study of $\LZ$ with the ESC08a(DD) and ESC08a(DI)
interactions, it was found that ESC08a(DD) fails to describe 
the observed excitation energy shift in $\LZ$, 
whereas ESC08a(DI) can describe 
a trend of the excitation energy shift but somewhat overestimates the experimental data. 
It suggests that a moderate density-dependence weaker than 
ESC08a(DD) may be favored.
In the hybrid version, we take the average of the ESC08a(DD) and ESC08a(DI) interactions as
\begin{eqnarray}
V^\textrm{e,o}_{\Lambda N}(r)&=&\frac{1}{2}\left[V^\textrm{e,o}_{\Lambda N}(k_f=k^\textrm{inp}; r)
+V^\textrm{e,o}_{\Lambda N}(k_f=\langle k_f \rangle_\Lambda,r)\right],
\nonumber\\
\end{eqnarray}
in which 
$\langle k_f \rangle_\Lambda$ is self-consistently determined for each state.
We label this interaction as ESC08a(Hyb). 

Since all of ESC08a(DD), ESC08a(DI), and  ESC08a(Hyb) are system-dependent interactions through the 
$k_f$ values determined for each system ($\LZ$),  these interactions 
reasonably reproduce the $\Lambda$ binding energies
$B_\La$ of $p$-shell hypernuclei because
the $\Lambda NG$ interactions are originally designed so as to reproduce the systematics of experimental 
$\Lambda$ binding energies in a wide mass number region. It should be noted that 
ESC08a(DI) is state-independent, but
ESC08a(DD) and  ESC08a(Hyb) are state-dependent interactions. Namely, 
ESC08a(DI), ESC08a(Hyb), and ESC08a(DD) have no, mild, and relatively strong density dependence of the $\Lambda NG$ interactions, respectively.

\subsection{Effective $\Lambda$-$\Lambda$ interactions}
For the $\Lambda$-$\Lambda$ interaction in $(0s)^2_\La$ states in $\LLZ$, 
we adopt the singlet-even part of the effective $\Lambda$-$\Lambda$ interactions used in Refs.~\cite{Hiyama:2002yj,Hiyama:2010zzd}, 
\begin{eqnarray}
V_\LLa(r)=\sum_{i=1,2,3} v^{1E}_i e^{-\gamma_i r^2},
\end{eqnarray}
with $\gamma_1=0.555$, $\gamma_2=1.656$, and $\gamma_1=8.163$ in fm$^{-2}$. 
The original values of the strength parameters $v^{1E}_{i}$ 
in the earlier work in Ref.~\cite{Hiyama:2002yj} are $v^{1E}_1=-10.96$, $v^{1E}_2=-141.75$, and 
$v^{1E}_3=2136.6$ in MeV, but a modified parameter $v^{1E}_3=2136.6a$ ($a=1.244$) 
was used in the later work in Ref.~\cite{Hiyama:2010zzd} to fit the revised experimental 
value of $\Delta B^\textrm{exp}_\LLa(\LLHe)=0.67\pm 0.17 $ MeV \cite{Nakazawa:2010zzb}.
By using $V_\LLa(r)$ with $a=1.244$ in the ESC08a(Hyb) calculation,  
we obtain $\Delta B_\LLa(\LLHe)=0.58$ MeV
for the frozen $(0s)^4$ $^4\textrm{He}$ core with the experimental size $R_N=1.455$ fm reduced from 
the charge radius data.
In the ESC08a(DD) and ESC08a(DI) calculations, we 
readjust the parameter as $a=1.51$ and $a=1.07$, respectively, which give almost the same $\Delta B_\LLa(\LLHe)$ values. 

\section{Results and Discussions} \label{sec:results}
\subsection{Properties of ground and excited states in $\LZ$}

\begin{table*}[ht]
\caption{\label{tab:L-gs}
Ground state properties of $\Lambda$ hypernuclei ($\LZ$). 
The $\Lambda$ distribution size ($r_\Lambda$), averaged Fermi momentum  ($\langle k_f \rangle_\Lambda$), core nuclear size ($R_N$), 
nuclear size change ($\delta_\Lambda(R_N)$),
nuclear energy change ($\delta_\Lambda(E_N)$), and 
the $\Lambda$ binding energy ($B_\Lambda$) in $\LZ$ are listed. 
The calculated results obtained with ESC08a(Hyb),  ESC08a(DI) and ESC08a(DD) are shown. 
The experimental $B_\Lambda$ values are taken from the data compilation in Ref.~\cite{Davis:2005mb}.
For $I\ne 0$ nuclei, spin-averaged values 
($\bar{B}_\Lambda$ (MeV))) 
of the experimental $\Lambda$ binding energies for spin doublet states are also 
shown. The spin-doublet splitting data are taken from Ref.~\cite{Tamura:2010zz} and references therein. 
The units of size, momentum, and energy values are fm,  fm$^{-1}$, and MeV, respectively.
}
\begin{center}
\begin{tabular}{cccccccccc}
\hline
\multicolumn{10}{c}{ESC08a(Hyb)}	\\
$\LZ(I^\pi)$ & $k^\textrm{inp}_f$		&	 $r_\Lambda$	&	$\langle k_f \rangle_\Lambda$	&	$R_N$	&	$\delta_\Lambda({R_N})$ &	
$\delta_\Lambda({E_N})$ 	&	$B_\Lambda$&$B^\textrm{exp}_{\Lambda}$	&	$\bar{B}^\textrm{exp}_{\Lambda}$	\\
$^{5}_\Lambda\textrm{He}(0^+_1)$	&	0.96	&	2.83 	&	0.97 	&	1.46 	&$	-	$&	$-$	&	3.53 	&	3.12(2)	&	$-$	\\
$^{7}_\Lambda\textrm{Li}(1^+_1)$	&	0.93	&	2.64 	&	0.93 	&	2.32 	&$	-0.24 	$&	0.23 	&	5.35 	&	5.58(3)	&	5.12(3)	\\
$^{8}_\Lambda\textrm{Li}(3/2^-_1)$	&	0.91	&	2.55 	&	0.96 	&	2.34 	&$	-0.15 	$&	0.26 	&	6.68 	&	6.80(3)	&		\\
$^{9}_\Lambda\textrm{Be}(0^+_1)$	&	0.90	&	2.59 	&	0.93 	&	2.57 	&$	-0.80 	$&	0.84 	&	6.53 	&	6.71(4)	&	$-$	\\
$^{10}_\Lambda\textrm{Be}(3/2^-_1)$	&	0.95	&	2.51 	&	0.99 	&	2.54 	&$	-0.18 	$&	0.34 	&	8.06 	&	9.11(22)	&		\\
%$^{10}_\Lambda\textrm{B}(3/2^-_1)$	&	0.94	&	2.51 	&	0.99 	&	2.56 	&$	-0.20 	$&	0.38 	&	8.01 	&	9.11(22)	&\\
$^{11}_\Lambda\textrm{Be}(0^+_1)$	&	1.04	&	2.47 	&	1.06 	&	2.39 	&$	-0.06 	$&	0.11 	&	9.01 	&	& $-$\\
$^{11}_\Lambda\textrm{B}(3^+_1)$	&	1.03	&	2.44 	&	1.08 	&	2.34 	&$	-0.05 	$&	0.08 	&	9.31 	&	10.24(5)	&	10.09(5)	\\
$^{12}_\Lambda\textrm{B}(3/2^-_1)$	&	1.07	&	2.40 	&	1.13 	&	2.29 	&$	-0.04 	$&	0.05 	&	10.06 	&	11.37(6)	&	11.27(6)	\\
$^{13}_\Lambda\textrm{C}(0^+_1)$	&	1.11	&	2.41 	&	1.15 	&	2.31 	&$	-0.04 	$&	0.05 	&	10.44 	&	11.69(12)	&	$-$ \\
\multicolumn{10}{c}{ESC08a(DI)}	\\
& $k^\textrm{inp}_f$		&	 $r_\Lambda$	&	$\langle k_f \rangle_\Lambda$	&	$R_N$	&	$\delta_\Lambda({R_N})$ &	
$\delta_\Lambda({E_N})$ 	&$B_\Lambda$ &	&	\\
$^{5}_\Lambda\textrm{He}(0^+_1)$	&	0.96	&	2.81 	&	0.97 	&	1.46 	&$	-	$&	$-$	&	3.60 	&	&	\\
$^{7}_\Lambda\textrm{Li}(1^+_1)$	&	0.93 	&	2.57 	&	0.95 	&	2.22 	&$	-0.33 	$&	0.59 	&	5.44 	&		&		\\
$^{8}_\Lambda\textrm{Li}(3/2^-_1)$	&	0.91	&	2.41 	&	1.00 	&	2.25 	&	$-0.24 	$&	0.79 	&	7.32 	&		&		\\
$^{9}_\Lambda\textrm{Be}(0^+_1)$	&	0.90 	&	2.44 	&	0.98 	&	2.44 	&$	-0.94 	$&	1.69 	&	7.04 	&		&		\\
$^{10}_\Lambda\textrm{Be}(3/2^-_1)$	&	0.95	&	2.39 	&	1.03 	&	2.43 	&$	-0.28 	$&	0.99 	&	8.69 	&	&		\\
%$^{10}_\Lambda\textrm{B}(3/2^-_1)$	&	0.94 	&	2.38 	&	1.03 	&	2.45 	&$	-0.31 	$&	1.10 	&	8.71 	&		&		\\
$^{11}_\Lambda\textrm{Be}(0^+_1)$	&	1.04 	&	2.41 	&	1.08 	&	2.34 	&$	-0.12 	$&	0.41 	&	9.32 	&		&		\\
$^{11}_\Lambda\textrm{B}(3^+_1)$	&	1.03 	&	2.36 	&	1.10 	&	2.29 	&$	-0.10 	$&	0.37 	&	9.97 	&		&		\\
$^{12}_\Lambda\textrm{B}(3/2^-_1)$	&	1.07 	&	2.33 	&	1.16 	&	2.24 	&$	-0.09 	$&	0.29 	&	10.94 	&		&		\\
$^{13}_\Lambda\textrm{C}(0^+_1)$	&	1.11 	&	2.35 	&	1.18 	&	2.26 	&$	-0.09 	$&	0.27 	&	11.07 	&		&		\\
\multicolumn{10}{c}{ESC08a(DD)}	\\
&		&	 $r_\Lambda$	&	$\langle k_f \rangle_\Lambda$	&	$R_N$	&	$\delta_\Lambda({R_N})$ &	
$\delta_\Lambda({E_N})$ 	&	$B_\Lambda$ &  & \\
$^{5}_\Lambda\textrm{He}(0^+_1)$	&		&	2.84 	&	0.96 	&	1.46 	&$	-	$&	$-$	&	3.49 	&		&	\\
$^{7}_\Lambda\textrm{Li}(1^+_1)$	&		&	2.66 	&	0.91 	&	2.40 	&$	-0.15 	$&	0.08 	&	5.43 	&		&		\\
$^{8}_\Lambda\textrm{Li}(3/2^-_1)$	&		&	2.61 	&	0.93 	&	2.42 	&$	-0.08 $	&	0.06 	&	6.43 	&	&		\\
$^{9}_\Lambda\textrm{Be}(0^+_1)$	&		&	2.67 	&	0.90 	&	2.69 	&$	-0.68 	$&	0.44 	&	6.43 	&		&		\\
$^{10}_\Lambda\textrm{Be}(3/2^-_1)$	&		&	2.57 	&	0.96 	&	2.62 	&$	-0.09 	$&	0.08 	&	7.84 	&		&		\\
%$^{10}_\Lambda\textrm{B}(3/2^-_1)$	&		&	2.58 	&	0.96 	&	2.65 	&$	-0.11 	$&	0.10 	&	7.76 	&		&		\\
$^{11}_\Lambda\textrm{Be}(0^+_1)$	&		&	2.49 	&	1.04 	&	2.44 	&$	-0.02 	$&	0.01 	&	8.92 	&		&		\\
$^{11}_\Lambda\textrm{B}(3^+_1)$	&		&	2.48 	&	1.06 	&	2.38 	&$	-0.01 	$&	0.00 	&	8.97 	&		&		\\
$^{12}_\Lambda\textrm{B}(3/2^-_1)$	&		&	2.45 	&	1.11 	&	2.33 	&	0.00 	&	0.00 	&	9.57 	&		&		\\
$^{13}_\Lambda\textrm{C}(0^+_1)$	&		&	2.44 	&	1.13 	&	2.35 	&	0.00 	&	0.00 	&	10.13 	&		&		\\
\hline		
\end{tabular}
\end{center}
\end{table*}

\begin{table*}[ht]
\caption{\label{tab:L-ex}
Properties of excited states $\LZ(I^\pi_n)$ in $\Lambda$ hypernuclei. 
The core nuclear size $R_{N}(\LZ)$ and size difference $R_{N}(\LZ)-R_{N,\textrm{gs}}(\LZ)$ 
from the ground state size in $\LZ$, the excitation energies $E_x$ in $^{A-2}Z$ and  $\LZ$ systems, 
and the excitation energy shift $\delta_\Lambda(E_x)$. 
The calculated values obtained  with ESC08a(Hyb) are shown together with  
$\delta_\Lambda(E_x)$ calculated with ESC08a(DI) and ESC08a(DD).
The units of sizes and energies are fm and MeV, respectively. For details of  the experimental data of excitation energies, see the captions of Figs.~\ref{fig:spe-yy1} and ~\ref{fig:spe-yy2}. 
}
\begin{center}
\begin{tabular}{cccccccccccc}
\hline
$\LZ(I^\pi)$&	$R_N(\LZ)$		&	 $R_{N}-R_{N,\textrm{gs}}$  & $E_x(^{A-2}Z)$ 	&	$E_x(^{A-2}Z)$	&	$E_x(\LZ)$&	$E_{x}(\LZ)$ 
& $\delta_\Lambda({E_x})$ &  $\delta_\Lambda({E_x})$ &  $\delta_\Lambda({E_x})$ &  $\delta_\Lambda({E_x})$\\
&	Hyb	&	Hyb	  & Hyb		&	exp	&	Hyb &	exp &Hyb & DI & DD & exp		\\
$^{7}_\Lambda\textrm{Li}(3^+_1)$	&	2.13 	&$	-0.19 	$&	2.08 	&	2.19 	&	1.50 	&	1.86 	&$	-0.58 	$&$	-1.19 	$&$	-0.21 	$&$	-0.33 	$\\
$^{8}_\Lambda\textrm{Li}(1/2^-_1)$	&	2.39 	&$	0.04 	$&	0.49 	&	0.48 	&	0.70 	&		&$	0.20 	$&$	0.38 	$&$	0.10 	$&$		$\\
$^{8}_\Lambda\textrm{Li}(7/2^-_1)$	&	2.25 	&$	-0.10 	$&	4.75 	&	4.63 	&	4.40 	&		&$	-0.34 	$&$	-0.92 	$&$	-0.05 	$&$		$\\
$^{9}_\Lambda\textrm{Be}(2^+_1)$	&	2.59 	&$	0.02 	$&	3.11 	&	3.04 	&	2.79 	&	3.04 	&$	-0.32 	$&$	-0.43 	$&$	-0.29 	$&$	0.00 	$\\
$^{10}_\Lambda\textrm{Be}(1/2^-_1)$	&	2.71 	&$	0.17 	$&	2.20 	&	2.78 	&	2.92 	&		&$	0.72 	$&$	1.23 	$&$	0.40 	$&$		$\\
$^{10}_\Lambda\textrm{Be}(5/2^-_1)$	&	2.55 	&$	0.01 	$&	2.02 	&	2.43 	&	2.13 	&		&$	0.10 	$&$	0.11 	$&$	0.07 	$&$		$\\
%$^{10}_\Lambda\textrm{B}(1/2^-_1)$	&	2.74 	&$	0.18 	$&	1.95 	&	2.78 	&	2.67 	&	0.00 	&$	0.73 	$&$	1.25 	$&$	0.40 	$&$		$\\
%$^{10}_\Lambda\textrm{B}(5/2^-_1)$	&	2.58 	&$	0.02 	$&	1.99 	&	2.43 	&	2.11 	&		&$	0.12 	$&$	0.13 	$&$	0.08 	$&$		$\\
$^{11}_\Lambda\textrm{Be}(2^+_1)$	&	2.37 	&$	-0.03 	$&	3.21 	&	3.37 	&	3.12 	&		&$	-0.10 	$&$	-0.25 	$&$	-0.01 	$&$		$\\
$^{11}_\Lambda\textrm{Be}(2^+_2)$	&	2.44 	&$	0.05 	$&	5.20 	&	5.96 	&	5.37 	&		&$	0.17 	$&$	0.38 	$&$	0.04 	$&$		$\\
$^{11}_\Lambda\textrm{B}(1^+_1)$	&	2.53 	&$	0.18 	$&	1.21 	&	0.72 	&	1.97 	&	1.67 	&$	0.77 	$&$	1.51 	$&$	0.23 	$&$	0.95 	$\\
$^{12}_\Lambda\textrm{B}(1/2^-_1)$	&	2.45 	&$	0.16 	$&	2.79 	&	2.13 	&	3.36 	&	3.00 	&$	0.57 	$&$	1.34 	$&$	0.02 	$&$	0.87 	$\\
$^{12}_\Lambda\textrm{B}(3/2^-_2)$	&	2.51 	&$	0.23 	$&	5.57 	&	5.02 	&	6.42 	&	6.02 	&$	0.85 	$&$	1.88 	$&$	0.11 	$&$	1.00 	$\\
$^{12}_\Lambda\textrm{B}(5/2^-_1)$	&	2.45 	&$	0.17 	$&	4.66 	&	4.45 	&	5.26 	&		&$	0.60 	$&$	1.39 	$&$	0.03 	$&$		$\\
$^{13}_\Lambda\textrm{C}(2^+_1)$	&	2.44 	&$	0.13 	$&	4.47 	&	4.44 	&	4.88 	&	4.89 	&$	0.41 	$&$	1.03 	$&$	-0.04 	$&$	0.45 	$\\
\hline		
\end{tabular}
\end{center}
\end{table*}

%%%%%%%%%%%%%%%%%%%%%%%%%%%%%%
\begin{figure}[htb]
\begin{center}
\includegraphics[width=8.0cm]{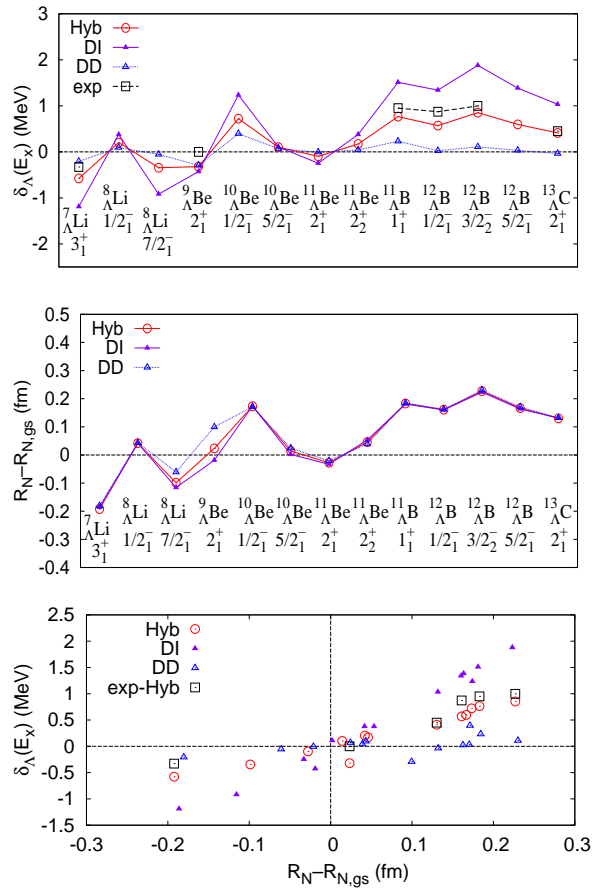} 	
\end{center}
%\vspace{0.5cm}
  \caption{(color online) 
(top) Excitation energy shift $\delta_\Lambda(E_x)$, 
(middle) nuclear size difference $(R_N-R_{N,\textrm{gs}})$, 
(bottom)  excitation energy shift plotted against the nuclear size difference. 
The results obtained with ESC08a(Hyb), ESC08a(DI), and ESC08a(DD) are shown. 
The experimental values of $\delta_\Lambda(E_x)$ are shown in the top panel, and those 
plotted against the size difference
calculated with ESC08a(Hyb) are shown in the bottom panel.
\label{fig:r-e-hybrid}}
\end{figure}
%%%%%%%%%%%%%%%%%%%%%%%%%%%%%

%%%%%%%%%%%%%%%%%%%%%%%%%%%%%%
\begin{figure}[!h]
\begin{center}
\includegraphics[width=8.5cm]{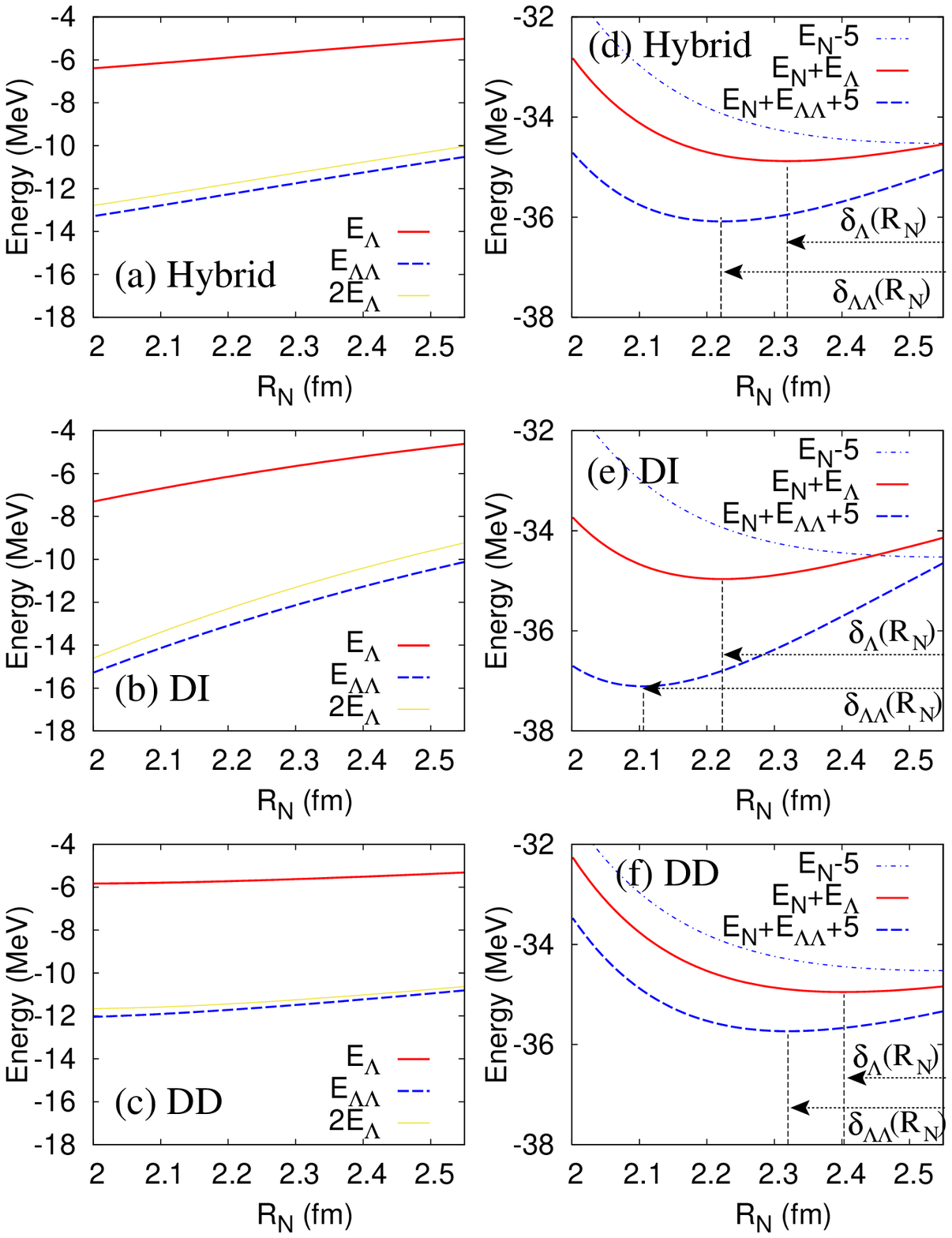} 	
\end{center}
%\vspace{0.5cm}
  \caption{(color online) 
$\epsilon$ dependence of energies for the $^{6}\textrm{Li}(1^+_1)$ core plotted as functions of the nuclear size $R_N(\epsilon)$.
Left : single-$\Lambda$ energy $(E_\Lambda(\epsilon))$ and  two-$\Lambda$ energy $(E_{\Lambda\Lambda}(\epsilon))$. 
$2 E_\Lambda(\epsilon)$, which correspond to the two-$\Lambda$ energy for the $V_{\Lambda\Lambda}=0$ case 
is also shown. 
Right: nuclear energy ($E_N(\epsilon)$), 
total energy $E(\epsilon;\LZ)=E_N(\epsilon)+E_\Lambda(\epsilon)$ of $\LZ$, and 
total energy $E(\epsilon;\LLZ)=E_N(\epsilon)+E_{\Lambda\Lambda}(\epsilon)$ of $\LLZ$. 
The values of $E_N(\epsilon)$ and $E(\epsilon;^A_\Lambda Z)$ are shifted by $-5$ MeV and $+5$ MeV, respectively. 
\label{fig:r-dep}}
\end{figure}
%%%%%%%%%%%%%%%%%%%%%%%%%%%%%

Structure properties of the ground states in $\LZ$ calculated with the 
ESC08a(Hyb), ESC08a(DI), and ESC08a(DD) interactions are listed in Table \ref{tab:L-gs} together with 
observed $\Lambda$ binding energies ($B^\textrm{exp}_{\La}$) and 
spin-averaged values ($\bar{B}^\textrm{exp}_{\La}$).
In the ESC08a(DI) result, significant core polarization by a $\Lambda$ particle in $\LZ$ generally occurs 
as seen in the nuclear size reduction $|\delta_\Lambda(R_N)|$. 
Compared with ESC08a(DI), the ESC08a(DD) result shows relatively small core polarization
because the density dependence of the $\Lambda NG$ interactions suppresses the nuclear size reduction.
The ESC08a(Hyb) calculation shows moderate core polarization between the ESC08a(DI) and ESC08a(DD) results. 
As explained in the previous section, 
the nuclear size in $\LZ$ is determined by the energy balance between the $\Lambda$ potential energy gain and 
the nuclear energy loss in the size reduction. As an example, we show the energy balance in $^7_\La \textrm{Li}$
in Fig.~\ref{fig:r-dep} showing $\epsilon$ dependence of energies plotted as functions of the core nuclear size 
$R_N(\epsilon)$. 
As $R_N(\epsilon)$ decreases, the $\La$ gains the potential energy through the $\Lambda$-$N$ interactions 
(see Figs.~\ref{fig:r-dep}(a)(b)(c)). 
The energy gain is 
largest in the ESC08a(DI) resulting in the largest size reduction among three calculations.
However, in the ESC08a(DD) result, the $\La$ potential energy gain 
is small because
the density dependence of the $\Lambda NG$ interactions  compensates the energy gain
and gives weak $R_N$ dependence of $E_\La(\LZ)$ (see Fig.~\ref{fig:r-dep}(c)).
Consequently,  the size reduction is suppressed in the ESC08a(DD) calculation.

The core polarization, {\it i.e.}, the nuclear size reduction causes the nuclear energy increase $\delta_\Lambda(E_N)$
which we call the core polarization energy. As shown in Table \ref{tab:L-gs},
$\delta_\Lambda(E_N)$ correlates sensitively with the size reduction.
ESC08a(DI),  ESC08a(Hyb), and ESC08a(DD) shows larger, moderate, and smaller core polarization energies 
$\delta_\Lambda(E_N)$.
In each calculation of ESC08a(DI),  ESC08a(Hyb), and ESC08a(DD), 
the relatively significant core polarization in $^7_\La\textrm{Li}$, $^8_\La\textrm{Li}$, $^9_\La\textrm{Be}$, and $^{10}_\La\textrm{Be}$ 
is found in the size and energy because of the remarkably developed 
$\alpha+d$, $\alpha+t$, $2\alpha$, and $2\alpha+n$ clustering
compared with those in $\LZ$ with $A>10$.

Tables \ref{tab:L-ex} shows energies for excited states in $\LZ$.
As discussed in the previous paper, the excitation energy shift $\delta_\La(E_x)$ by a 
$\La$ in $\LZ$ can be qualitatively described by the size difference 
$R_N-R_{N,\textrm{gs}}$ between the ground and excited states.
Note that the core polarization gives only a minor effect to the size difference 
$R_N-R_{N,\textrm{gs}}$ and also the excitation energy shift $\delta_\La(E_x)$. It means that
the excitation energy shift in $\LZ$ dominantly originates in the original size difference between the ground and
excited states in $^{A-2}Z$. 
Since the $\Lambda$ binding is generally deeper in higher nuclear density,  
a smaller size state tends to gain more $\Lambda$ binding 
than a larger size state. As a result, $\delta_\La(E_x)$ and $R_N-R_{N,\textrm{gs}}$ show the positive correlation.

To  see the correlation 
and its interaction dependence more quantitatively,
 $\delta_\La(E_x)$ and $R_N-R_{N,\textrm{gs}}$ are plotted in Fig.~\ref{fig:r-e-hybrid}. 
The ESC08a(DD) result shows no or only slight excitation energy shift
because the density dependence of the $\Lambda NG$ interactions cancels 
positive correlation 
between $\delta_\La(E_x)$ and $R_N-R_{N,\textrm{gs}}$. The ESC08a(DD) result is inconsistent with the experimental
$\delta_\La(E_x)$ values. 
The ESC08a(DI) calculation for the density-independent $\Lambda NG$ interactions 
shows the strong correlation, {\it i.e.}, the large excitation energy shift, and 
describes the qualitative trend of the experimental data of $\delta_\La(E_x)$. However, it quantitatively overestimates the 
data. The ESC08a(Hyb) interaction for the mild density dependence
gives moderate correlation and reasonably reproduces the systematics of experimental 
$\delta_\La(E_x)$ in $p$-shell $\LZ$. 
%Hereafter, we mainly discuss results obtained by the ESC08a(Hyb). 

\subsection{Ground state properties of $\LLZ$}
Calculated results of ground state properties of $\LLZ$ are shown in Table \ref{tab:LL-gs}.
Compared with $\LZ$, further nuclear size reduction occurs, especially, in $\LLZ$
with $A\le 10$.   The core nuclear size in $\LLZ$ is determined so as to optimize 
the potential energies of two $\Lambda$s and nuclear energy loss in the size reduction. 
The energy balance in $^7_\La \textrm{Li}$ and 
$^{\ \ 8}_\LLa \textrm{Li}$ is illustrated in Fig.~\ref{fig:r-dep}, in which energies are plotted 
as functions of $R_N(\epsilon)$. As shown in the figure, 
the $R_N$ dependence of two-$\La$ energy $E_\LLa$ is almost same as that of 
$2E_\La$ (twice of single-$\La$ energy)  with a constant shift. It means that the $\La$-$\La$ interaction contribution 
is negligible in the $R_N$ dependence. 
As seen in Table \ref{tab:LL-gs}, 
among three calculations, the ESC08a(DI) (ESC08a(DD)) calculation generally shows the
strong (weak) $R_N$ dependence of the $\La$ potential energy
and the larger (smaller) size reduction than other interactions.
The size reduction, {\it i.e.}, the core polarization 
is moderate in the ESC08a(Hyb) case. More detailed discussion of the size reduction
is given later. 

Let us focus on the $\LLa$ binding in $\LLZ$. 
The calculated values of the $\Lambda$-$\Lambda$ binding energy $\Delta B_{\Lambda\Lambda}$ show
strong interaction dependence.
In general, values of $\Delta B_{\Lambda\Lambda}$ obtained with 
ESC08a(Hyb) are moderate among three calculations, whereas those with ESC08a(DI) and 
ESC08a(DD) are larger and smaller than the ESC08a(Hyb) result, respectively. 

Moreover, each calculation shows rather strong
system (mass-number) dependence of $\Delta B_{\Lambda\Lambda}$.
In contrast, the values obtained without the core polarization 
$\Delta B^\textrm{w/o cp}_{\Lambda\Lambda}$ (the $\Lambda$-$\Lambda$ binding energy calculated with frozen cores)
show weak system dependence. 
Furthermore, the $\LLa$ bond energy 
${\cal V}^\textrm{bond}_\LLa$ given in \eqref{eq:vbond} is almost system independent  except for 
$\LLHe$.

\begin{figure}[!h]
\begin{center}
\includegraphics[width=8.0cm]{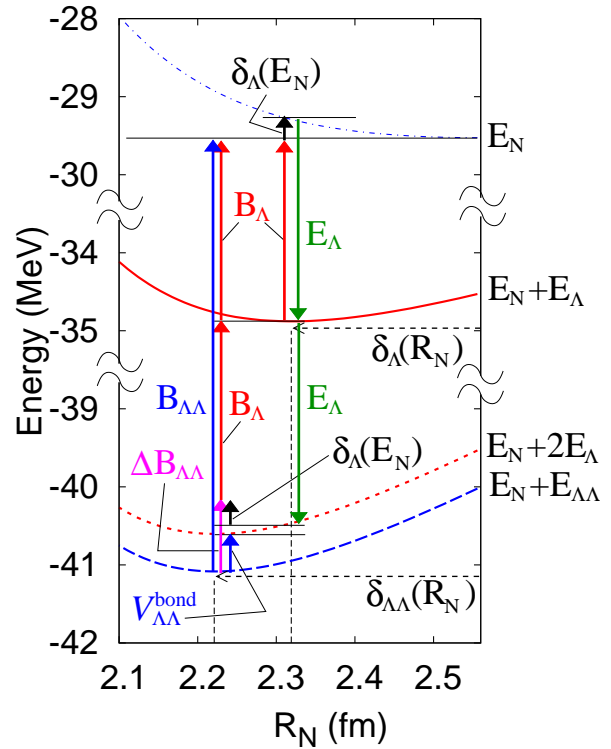} 	
\end{center}
%\vspace{0.5cm}
  \caption{(color online) 
Energy counting in $^{7}_\La\textrm{Li}(1^+_1)$ and $^{8}_\LLa\textrm{Li}(1^+_1)$
 for the $^{6}\textrm{Li}(1^+_1)$ core.
$R_N(\epsilon)$ dependences of 
$E_N$, $E_N+E_\La$, $E_N+E_\LLa$, and $E_N+2E_\La$ calculated with
ESC08a(Hyb) are shown by dash-dotted, solid, dashed, and dotted lines, 
respectively. 
\label{fig:e-count}}
\end{figure}

The main origin of the deviation of $\Delta B_\LLa$ from ${\cal V}^\textrm{bond}_\LLa$
is the core polarization energy by a $\Lambda$. 
As discussed previously, 
a $\Lambda$ particle in $\LZ$ reduces the core nuclear size to gain its potential energy. The size reduction
induces the core polarization energy, {\it i.e.}, the nuclear energy increase $\delta_\Lambda (E_N)$.
Provided that the core polarization, namely, the nuclear density $\rho_N(r)$, is same in $\LZ$ and $\LLZ$, following relations are
obtained using \eqref{eq:BL}:
\begin{eqnarray}
B_\LLa(V_\LLa=0)&=&-\left[ 2E_\Lambda+\delta_\Lambda(E_N)\right]\\
B_\LLa&=&{\cal V}^\textrm{bond}_\LLa+B_\LLa(V_\LLa=0)\nonumber\\
&= &-\delta_\La(E_N)-2 E_\La+{\cal V}^\textrm{bond}_\LLa, \\
\Delta B_\LLa&=&{\cal V}^\textrm{bond}_\LLa + \delta_\La(E_N). 
\end{eqnarray}
The essential point is that the single-$\Lambda$ energy $E_\La$ appears twice 
but the core polarization energy $\delta_\La(E_N)$ does only once in $B_\LLa$. 
As an example, the energy counting in $^{\ \ 8}_\LLa\textrm{Li}$
is shown in Fig.~\ref{fig:e-count}. 
At each $R_N(\epsilon)$ for a fixed $\epsilon$,  the above 
relations are exactly satisfied and 
the difference $\Delta B_\LLa-{\cal V}^\textrm{bond}_\LLa$ is simply given by
$\delta_\La(E_N)$. 
In other words, the deviation of $\Delta B_\LLa$ from ${\cal V}^\textrm{bond}_\LLa+\delta_\La(E_N)$
comes from the additional core polarization effect
from the second $\La$ particle. Comparing the energies at $R_N=2.32$ fm for $^{7}_\La\textrm{Li}$
and those at $R_N=2.22$ fm for $^{\ \ 8}_\LLa\textrm{Li}$, the additional core polarization effect
in $\Delta B_\LLa$ is found to be relatively minor. 

\begin{figure}[!h]
\begin{center}
\includegraphics[width=8.0cm]{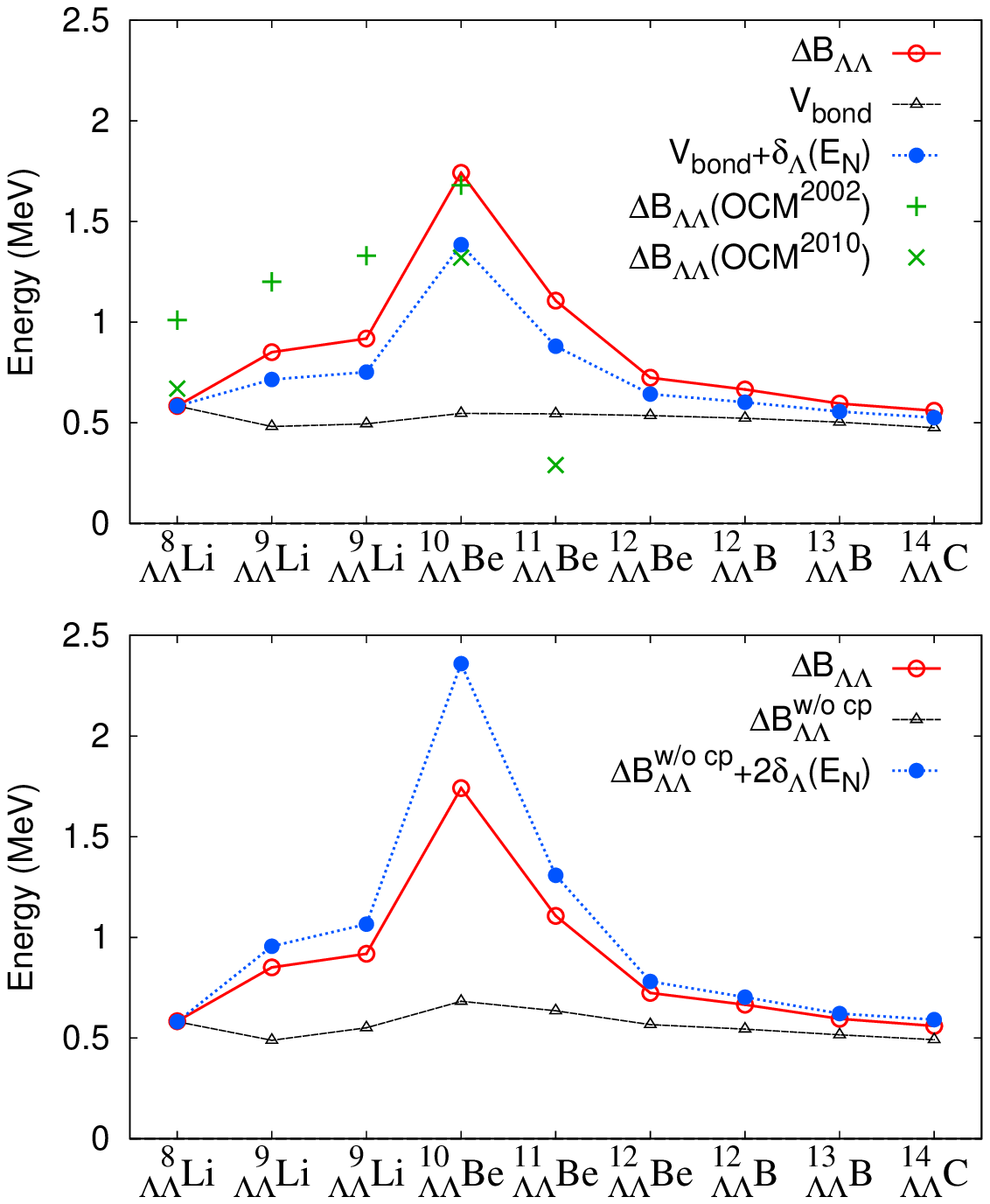} 	
\end{center}
%\vspace{0.5cm}
  \caption{(color online) 
Upper: the values of 
$\Delta B_{\Lambda\Lambda}$, ${\cal V}^\textrm{bond}_{\Lambda\Lambda}$, and 
${\cal V}^\textrm{bond}_{\Lambda\Lambda}+\delta_\Lambda(E_N)$ for $\LLZ$
calculated with ESC08a(Hyb).
($\delta_\Lambda(E_N)$
is the nuclear energy change $\delta_\Lambda(E_N)$ in $\LZ$.) 
For comparison,
the theoretical $\Delta \bar B_{\Lambda\Lambda}$ (spin-averaged values) 
of the OCM cluster model calculation
in Ref.~\cite{Hiyama:2002yj} (OCM$^{2002}$) and those in Ref.~\cite{Hiyama:2010zzd} (OCM$^{2010}$) 
are also shown. 
Lower: $\Delta B_{\Lambda\Lambda}$, $\Delta B^\textrm{w/o cp}_{\Lambda\Lambda}$ , and 
$\Delta B^\textrm{w/o cp}_{\Lambda\Lambda}+2\delta_\Lambda(E_N)$ for $\LLZ$.
\label{fig:vbond}}
\end{figure}

In the upper panel of Fig.~\ref{fig:vbond}, the calculated values of 
$\Delta B_\LLa$, ${\cal V}^\textrm{bond}_\LLa$, 
and ${\cal V}^\textrm{bond}_\LLa + \delta_\La(E_N)$ obtained with ESC08a(Hyb) are compared.
As seen in the figure, ${\cal V}^\textrm{bond}_\LLa$ is nearly system independent. 
In $A\ge 12$ systems having minor core polarization, 
$\Delta B_\LLa$ is almost consistent with ${\cal V}^\textrm{bond}_\LLa$
and shows system independent behavior. 
However, in lighter-mass nuclei, 
$^{\ \ 8}_\LLa\textrm{Li}$, $^{\  \ 9}_\LLa\textrm{Li}$, $^{10}_\LLa\textrm{Be}$, and $^{11}_\La\textrm{Be}$, 
the departure of $\Delta B_\LLa$ from ${\cal V}^\textrm{bond}_\LLa$ is significantly large
because of the remarkable core polarization in the developed clustering. 
As is expected from the above-mentioned discussion, the approximation
$\Delta B_\LLa\approx {\cal V}^\textrm{bond}_\LLa + \delta_\La(E_N)$ is roughly satisfied 
indicating that 
the system dependence of $\Delta B_\LLa$  in the $A\le 10$ region is mainly described by $\delta_\La(E_N)$. In other words, 
the strong system dependence of $\Delta B_\LLa$  in the $A\le 10$ region originates in the 
significant core polarization energy because of developed clustering. 
The result suggests that,  in order to extract clean information of the $\Lambda$-$\Lambda$ binding 
from observations of binding energies of double-$\Lambda$ hypernuclei,
heavier-mass nuclei are favored because they are more 
free from the core polarization effect rather than very light-mass ones.

The system dependence of the $\La$-$\La$ binding energies in the $A\le 10$ region 
has been discussed in detail with the OCM cluster model calculation in Ref.~\cite{Hiyama:2002yj}.
In Fig.~\ref{fig:vbond}, 
the theoretical $\Delta \bar B_\LLa$ (spin-averaged values) of the OCM cluster model calculations 
are also shown for comparison. 
The $A$-dependent behavior of $\Delta B_\LLa$ in the present calculation 
is similar to the result of Ref.~\cite{Hiyama:2002yj}. 
The system-independent 
behavior of ${\cal V}^\textrm{bond}_\LLa$ in the present result is also consistent 
with their result.
It should be commented that, in the earlier work of OCM cluster model calculations in Ref.~\cite{Hiyama:2002yj},  
the $\La$-$\La$ interaction was adjusted to an old data of the $^{\ \ 6}_{\LLa}\textrm{He}$
binding energy. It is slightly stronger than that used in the later work in Ref.~\cite{Hiyama:2010zzd}, which 
was adjusted to the revised data.

In the case that the core polarization effect is small enough and
can be treated perturbatively, $\Delta B_\LLa$ is given simply using 
$\delta_\La(E_N)$ (core polarization energy) and $\Delta B^\textrm{w/o cp}_\LLa$
(the $\La$-$\La$ binding energy for the frozen core)
as,
\begin{equation}
\Delta B_\LLa=\Delta B^\textrm{w/o cp}_\LLa+2\delta_\La(E_N), 
\end{equation}
as discussed in Ref.~\cite{Lanskoy:1997xq}. As shown in the lower panel of 
Fig.~\ref{fig:vbond}, the relation is not fulfilled in lighter-mass nuclei, in which 
the perturbative evaluation is too simple to quantitatively describe 
the core polarization in the remarkable clustering.

Comparing with the ESC08a(Hyb) result, the ESC08a(DI) calculation 
gives much larger $\Delta B_\LLa$ values by a factor of $\sim 2$
because of larger core polarization 
energy $\delta_\La(E_N)$, whereas the 
ESC08a(DD) result shows generally small $\Delta B_\LLa$
because the core polarization is suppressed by
the density dependence of the $\La NG$ interactions.

As shown later, the calculated $\Delta B_\LLa$ and ${\cal V}^\textrm{bond}_\LLa$ values for excited states are 
similar to those for the ground state. The present result of $\Delta B_\LLa$ for $\LLBe(2^+_1)$ 
obtained with ESC08a(Hyb) is consistent  with the experimental observation and in reasonable agreement with the 
theoretical value of the OCM cluster model calculation \cite{Hiyama:2010zzd}. However, 
the ESC08a(DI) and ESC08a(DD) results much overestimate and underestimate the observed 
$\Delta B_\LLa(\LLBe(2^+_1))$, respectively.
Although the ESC08a(DI) and ESC08a(DD) calculations for other double-$\Lambda$ hypernuclei 
are not excluded by experimental observations, the  ESC08a(Hyb) result is likely to be favored. 

\begin{table*}[ht]
\caption{\label{tab:LL-gs}
Ground state properties of double-$\Lambda$ hypernuclei. 
The $\Lambda$ distribution size ($r_\Lambda$), averaged Fermi momentum  ($\langle k_f \rangle_\Lambda$), core nuclear size ($R_N$), 
nuclear size change ($\delta_{\Lambda\Lambda}(R_N)$),
nuclear energy change ($\delta_{\Lambda\Lambda}(E_N)$), 
two-$\La$ binding energy  ($B_{\Lambda\Lambda}$), 
$\La$-$\La$ binding energy ($\Delta B_{\Lambda\Lambda}$), 
and $\Lambda\Lambda$ bond energy 
(${\cal V}^\textrm{bond}_{\Lambda\Lambda}$) in $\LLZ$. 
The $\La$-$\La$ binding energy calculated without the core polarization ($\Delta B^\textrm{w/o cp}_{\Lambda\Lambda}$)
is also shown.
The units of size, momentum, and energy values are fm, fm$^{-1}$, and MeV,
respectively.
}
\begin{center}
\begin{tabular}{cccccccccc}
\hline
\multicolumn{10}{c}{Hybrid}	\\
&	 $r_\Lambda$	&	$\langle k_f \rangle_\Lambda$	&	$R_N$	&	$\delta_{\Lambda\Lambda}({R_N})$ &	
$\delta_{\Lambda\Lambda}({E_N})$ 	&	$B_{\Lambda\Lambda}$&$\Delta B_{\Lambda\Lambda}$	&	${\cal V}^\textrm{bond}_{\Lambda\Lambda}$  &$\Delta B^\textrm{w/o cp}_{\Lambda\Lambda}$  \\
$^{6}_{\Lambda\Lambda}\textrm{He}(0^+_1)$	&	2.81 	&	0.93 	&	1.46 	&$	-	$&$	-	$&$	7.64 	$&	0.58 	&	0.58 	&	0.58 	\\
$^{8}_{\Lambda\Lambda}\textrm{Li}(1^+_1)$	&	2.62 	&	0.93 	&	2.22 	&$	-0.33 	$&$	0.60 	$&$	11.56 	$&	0.85 	&	0.48 	&	0.49 	\\
$^{9}_{\Lambda\Lambda}\textrm{Li}(3/2^-_1)$	&	2.53 	&	0.97 	&	2.27 	&$	-0.23 	$&$	0.68 	$&$	14.27 	$&	0.92 	&	0.49 	&	0.55 	\\
$^{10}_{\Lambda\Lambda}\textrm{Be}(0^+_1)$	&	2.53 	&	0.96 	&	2.44 	&$	-0.93 	$&$	1.68 	$&$	14.80 	$&	1.74 	&	0.55 	&	0.68 	\\
$^{11}_{\Lambda\Lambda}\textrm{Be}(3/2^-_1)$	&	2.49 	&	1.01 	&	2.44 	&$	-0.27 	$&$	0.89 	$&$	17.22 	$&	1.11 	&	0.54 	&	0.64 	\\
%$^{11}_{\Lambda\Lambda}\textrm{B}(3/2^-_1)$	&	2.48 	&	1.01 	&	2.46 	&$	-0.30 	$&$	1.00 	$&$	17.21 	$&	1.18 	&	0.55 	&	0.65 	\\
$^{12}_{\Lambda\Lambda}\textrm{Be}(0^+_1)$	&	2.47 	&	1.06 	&	2.35 	&$	-0.10 	$&$	0.31 	$&$	18.74 	$&	0.72 	&	0.54 	&	0.57 	\\
$^{12}_{\Lambda\Lambda}\textrm{B}(3^+_1)$	&	2.45 	&	1.08 	&	2.31 	&$	-0.08 	$&$	0.24 	$&$	19.29 	$&	0.67 	&	0.52 	&	0.54 	\\
$^{13}_{\Lambda\Lambda}\textrm{B}(3/2^-_1)$	&	2.42 	&	1.13 	&	2.26 	&$	-0.07 	$&$	0.15 	$&$	20.71 	$&	0.60 	&	0.50 	&	0.52 	\\
$^{14}_{\Lambda\Lambda}\textrm{C}(0^+_1)$	&	2.43 	&	1.15 	&	2.28 	&$	-0.07 	$&$	0.14 	$&$	21.44 	$&	0.56 	&	0.48 	&	0.49 	\\
\multicolumn{10}{c}{DI}	\\
&	 $r_\Lambda$	&	$\langle k_f \rangle_\Lambda$	&	$R_N$	&	$\delta_\LLa({R_N})$ &	
$\delta_\LLa({E_N})$ 	&	$B_{\Lambda\Lambda}$&$\Delta B_{\Lambda\Lambda}$	&	${\cal V}^\textrm{bond}_{\Lambda\Lambda}$  &$\Delta B^\textrm{w/o cp}_{\Lambda\Lambda}$ \\
$^{6}_{\Lambda\Lambda}\textrm{He}(0^+_1)$	&	2.79 	&	0.94 	&	1.46 	&$	-	$&$	-	$&$	7.78 	$&	0.58 	&	0.58 	&	0.58 	\\
$^{8}_{\Lambda\Lambda}\textrm{Li}(1^+_1)$	&	2.48 	&	0.98 	&	2.11 	&$	-0.45 	$&$	1.49 	$&$	12.58 	$&	1.70 	&	0.74 	&	0.89 	\\
$^{9}_{\Lambda\Lambda}\textrm{Li}(3/2^-_1)$	&	2.33 	&	1.03 	&	2.14 	&$	-0.35 	$&$	2.10 	$&$	16.73 	$&	2.10 	&	0.75 	&	1.02 	\\
$^{10}_{\Lambda\Lambda}\textrm{Be}(0^+_1)$	&	2.31 	&	1.04 	&	2.28 	&$	-1.10 	$&$	3.83 	$&$	17.57 	$&	3.49 	&	0.89 	&	1.03 	\\
$^{11}_{\Lambda\Lambda}\textrm{Be}(3/2^-_1)$	&	2.30 	&	1.08 	&	2.30 	&$	-0.42 	$&$	2.69 	$&$	19.98 	$&	2.61 	&	0.91 	&	1.15 	\\
%$^{11}_{\Lambda\Lambda}\textrm{B}(3/2^-_1)$	&	2.29 	&	1.08 	&	2.30 	&$	-0.46 	$&$	2.97 	$&$	20.20 	$&	2.78 	&	0.90 	&	1.16 	\\
$^{12}_{\Lambda\Lambda}\textrm{Be}(0^+_1)$	&	2.37 	&	1.11 	&	2.27 	&$	-0.19 	$&$	1.18 	$&$	20.32 	$&	1.68 	&	0.96 	&	1.07 	\\
$^{12}_{\Lambda\Lambda}\textrm{B}(3^+_1)$	&	2.33 	&	1.12 	&	2.23 	&$	-0.16 	$&$	1.09 	$&$	21.53 	$&	1.59 	&	0.93 	&	1.04 	\\
$^{13}_{\Lambda\Lambda}\textrm{B}(3/2^-_1)$	&	2.32 	&	1.17 	&	2.20 	&$	-0.13 	$&$	0.75 	$&$	23.26 	$&	1.38 	&	0.90 	&	0.99 	\\
$^{14}_{\Lambda\Lambda}\textrm{C}(0^+_1)$	&	2.34 	&	1.19 	&	2.22 	&$	-0.13 	$&$	0.64 	$&$	23.46 	$&	1.32 	&	0.90 	&	0.98 	\\
\multicolumn{10}{c}{DD}	\\
&	 $r_\Lambda$	&	$\langle k_f \rangle_\Lambda$	&	$R_N$	&	$\delta_\LLa({R_N})$ &	
$\delta_\LLa({E_N})$ 	&	$B_{\Lambda\Lambda}$&$\Delta B_{\Lambda\Lambda}$	&	${\cal V}^\textrm{bond}_{\Lambda\Lambda}$  &$\Delta B^\textrm{w/o cp}_{\Lambda\Lambda}$ \\
$^{6}_{\Lambda\Lambda}\textrm{He}(0^+_1)$	&	2.83 	&	0.92 	&	1.46 	&$	-	$&$	-	$&$	7.57 	$&	0.58 	&	0.58 	&	0.58 	\\
$^{8}_{\Lambda\Lambda}\textrm{Li}(1^+_1)$	&	2.69 	&	0.90 	&	2.32 	&$	-0.24 	$&$	0.23 	$&$	11.21 	$&	0.36 	&	0.23 	&	0.18 	\\
$^{9}_{\Lambda\Lambda}\textrm{Li}(3/2^-_1)$	&	2.64 	&	0.93 	&	2.37 	&$	-0.13 	$&$	0.18 	$&$	13.17 	$&	0.30 	&	0.20 	&	0.18 	\\
$^{10}_{\Lambda\Lambda}\textrm{Be}(0^+_1)$	&	2.67 	&	0.91 	&	2.58 	&$	-0.79 	$&$	0.79 	$&$	13.64 	$&	0.77 	&	0.19 	&	0.31 	\\
$^{11}_{\Lambda\Lambda}\textrm{Be}(3/2^-_1)$	&	2.60 	&	0.97 	&	2.57 	&$	-0.15 	$&$	0.22 	$&$	16.00 	$&	0.33 	&	0.19 	&	0.21 	\\
%$^{11}_{\Lambda\Lambda}\textrm{B}(3/2^-_1)$	&	2.61 	&	0.96 	&	2.59 	&$	-0.17 	$&$	0.26 	$&$	15.88 	$&	0.36 	&	0.19 	&	0.21 	\\
$^{12}_{\Lambda\Lambda}\textrm{Be}(0^+_1)$	&	2.54 	&	1.03 	&	2.42 	&$	-0.03 	$&$	0.03 	$&$	18.04 	$&	0.20 	&	0.19 	&	0.18 	\\
$^{12}_{\Lambda\Lambda}\textrm{B}(3^+_1)$	&	2.53 	&	1.05 	&	2.37 	&$	-0.02 	$&$	0.01 	$&$	18.13 	$&	0.20 	&	0.19 	&	0.19 	\\
$^{13}_{\Lambda\Lambda}\textrm{B}(3/2^-_1)$	&	2.50 	&	1.10 	&	2.33 	&$	0.00 	$&$	0.00 	$&$	19.34 	$&	0.21 	&	0.32 	&	0.21 	\\
$^{14}_{\Lambda\Lambda}\textrm{C}(0^+_1)$	&	2.49 	&	1.12 	&	2.35 	&$	0.00 	$&$	0.00 	$&$	20.41 	$&	0.16 	&	0.33 	&	0.16 	\\
\hline		
\end{tabular}
\end{center}
\end{table*}

%\subsection{$\La$-$\La$ binding in $\LLZ$}

%%%%%%%%%%%%%%%%%%%%%%%%%%%%%%
%%%%%%%%%%%%%%%%%%%%%%%%%%%%%

%%%%%%%%%%%%%%%%%%%%%%%%%%%%%%
%%%%%%%%%%%%%%%%%%%%%%%%%%%%%

\subsection{Properties of excited states of $\LLZ$}

\begin{table}[ht]
\caption{\label{tab:LL-ex}
$\La$-$\La$ binding energy $\Delta B_{\Lambda\Lambda}$
 (MeV) and $\Lambda\Lambda$ bond energy ${\cal V}^\textrm{bond}_{\Lambda\Lambda}$ (MeV) for excited states in $\LLZ$
calculated with ESC08a(Hyb), ESC08a(DI), and ESC08a(DD).
}
\begin{center}
\begin{tabular}{ccccccc}
\hline
$\LLZ(J^\pi)$ &$\Delta{B}_{\Lambda\Lambda}$ & ${\cal V}^\textrm{bond}_{\Lambda\Lambda}$ &
$\Delta{B}_{\Lambda\Lambda}$ & ${\cal V}^\textrm{bond}_{\Lambda\Lambda}$ &
$\Delta{B}_{\Lambda\Lambda}$ & ${\cal V}^\textrm{bond}_{\Lambda\Lambda}$  \\
	& Hyb &Hyb& DI&DI&DD&DD\\
$^{8}_{\Lambda\Lambda}\textrm{Li}(3^+_1)$	&	0.77 	&	0.49 	&	1.40 	&	0.63 	&	0.40 	&	0.32 	\\
$^{9}_{\Lambda\Lambda}\textrm{Li}(1/2^-_1)$	&	0.98 	&	0.50 	&	2.19 	&	0.79 	&	0.32 	&	0.19 	\\
$^{9}_{\Lambda\Lambda}\textrm{Li}(7/2^-_1)$	&	0.96 	&	0.48 	&	2.01 	&	0.63 	&	0.36 	&	0.24 	\\
$^{10}_{\Lambda\Lambda}\textrm{Be}(2^+_1)$	&	1.49 	&	0.53 	&	3.34 	&	0.85 	&	0.44 	&	0.19  \\
$^{11}_{\Lambda\Lambda}\textrm{Be}(1/2^-_1)$	&	1.29 	&	0.59 	&	2.77 	&	1.01 	&	0.45 	&	0.21 	\\
$^{11}_{\Lambda\Lambda}\textrm{Be}(5/2^-_1)$	&	1.19 	&	0.54 	&	2.77 	&	0.90 	&	0.36 	&	0.19 	\\
%$^{11}_{\Lambda\Lambda}\textrm{B}(1/2^-_1)$	&	1.34 	&	0.59 	&	2.91 	&	1.01 	&	0.48 	&	0.21 	\\
%$^{11}_{\Lambda\Lambda}\textrm{B}(5/2^-_1)$	&	1.27 	&	0.55 	&	2.96 	&	0.89 	&	0.40 	&	0.20 	\\
$^{12}_{\Lambda\Lambda}\textrm{Be}(2^+_1)$	&	0.72 	&	0.53 	&	1.67 	&	0.94 	&	0.21 	&	0.20 	\\
$^{12}_{\Lambda\Lambda}\textrm{Be}(2^+_2)$	&	0.74 	&	0.55 	&	1.67 	&	1.00 	&	0.21 	&	0.18 	\\
$^{12}_{\Lambda\Lambda}\textrm{B}(1^+_1)$	&	0.84 	&	0.57 	&	1.88 	&	1.03 	&	0.23 	&	0.18 	\\
$^{13}_{\Lambda\Lambda}\textrm{B}(1/2^-_1)$	&	0.71 	&	0.54 	&	1.66 	&	1.00 	&	0.17 	&	0.16 	\\
$^{13}_{\Lambda\Lambda}\textrm{B}(3/2^-_2)$	&	0.79 	&	0.56 	&	1.80 	&	1.03 	&	0.19 	&	0.16 	\\
$^{13}_{\Lambda\Lambda}\textrm{B}(5/2^-_1)$	&	0.73 	&	0.54 	&	1.69 	&	1.00 	&	0.17 	&	0.16 	\\
$^{14}_{\Lambda\Lambda}\textrm{C}(2^+_1)$	&	0.65 	&	0.50 	&	1.54 	&	0.98 	&	0.12 	&	0.12 	\\
\hline		
\end{tabular}
\end{center}
\end{table}

%%%%%%%%%%%%%%%%%%%%%%%%%%%%%%
\begin{figure}[!h]
\begin{center}
\includegraphics[width=6.0cm]{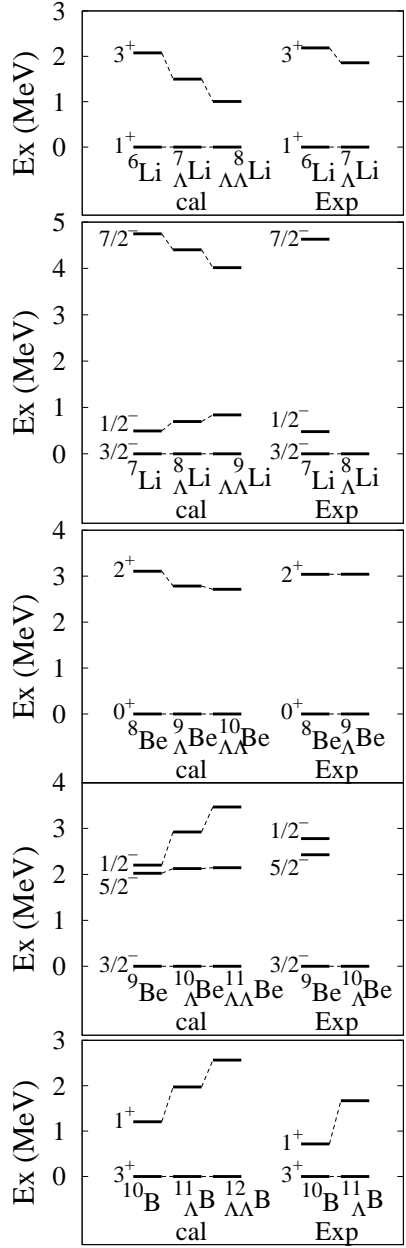} 	
\end{center}
%\vspace{0.5cm}
  \caption{
Energy spectra calculated with 
ESC08a(Hyb) and the experimental spectra.
The experimental data for $^{A-2}Z$ are taken from Refs.~\cite{Tilley:2002vg,Tilley:2004zz},
and those for  $\LZ$ are 
from Refs.~\cite{Tamura:2010zz,Tanida:2000zs,Ukai:2006zp,Akikawa:2002tm,Ajimura:2001na}.
The excitation energy of $^8\textrm{Be}(2^+)$ is calculated with the resonance energy obtained by the resonating group method (RGM). 
The experimental data for $\LZ$ are the spin-averaged values reduced from  the excitation energies of spin doublet states.
\label{fig:spe-yy1}}
\end{figure}
%%%%%%%%%%%%%%%%%%%%%%%%%%%%%

%%%%%%%%%%%%%%%%%%%%%%%%%%%%%%
\begin{figure*}[!h]
\begin{center}
\includegraphics[width=16.0cm]{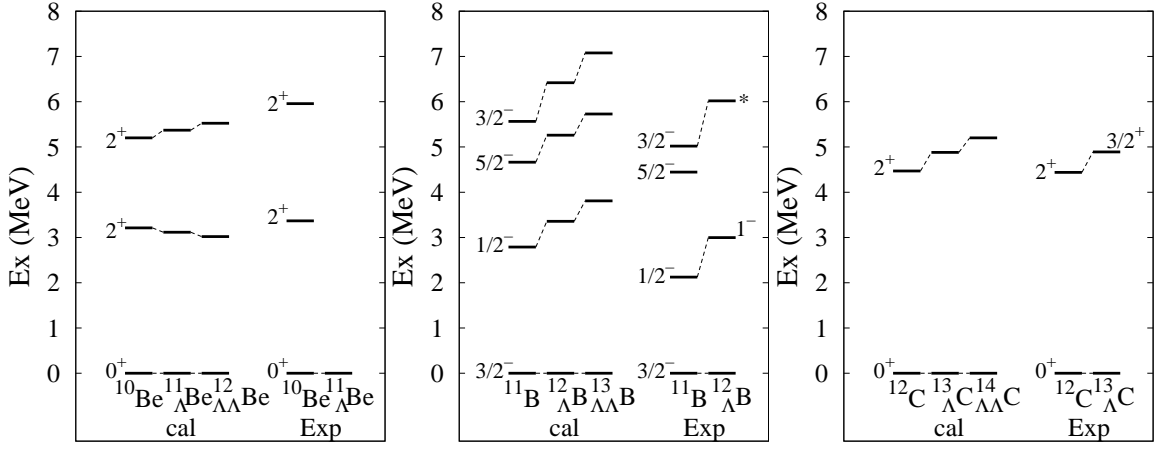} 	
\end{center}
%\vspace{0.5cm}
  \caption{Energy spectra calculated with 
ESC08a(Hyb) and experimental spectra.
The experimental data for $^{A-2}Z$ are taken from Refs.~\cite{Tilley:2004zz,Kelley:2012qua,AjzenbergSelove:1990zh},
and those for  $\LZ$ are 
from Refs.~\cite{Tamura:2010zz,Ajimura:2001na,Miura:2005mh,Ma:2010zzb,Kohri:2001nc,Tang:2014atx,Hosomi:2015fma}.
For the experimental spectra of $^{12}_\Lambda\textrm{B}(1/2^-_1)$
and  $^{13}_\Lambda\textrm{C}(2^+)$, 
non-spin-averaged values $E_x(1^-)$ and $E_x(3/2^+)$  are used, respectively.
For $^{12}_\Lambda\textrm{B}(3/2^-_2)$, the experimental values of
 $E_x(1^-)$ in  $^{12}_\Lambda\textrm{B}$ and $E_x(2^-)$ of the mirror state in $^{12}_\Lambda\textrm{C}$ are averaged
by assuming the same Coulomb shift in $^{12}_\Lambda\textrm{B}(3/2^-_2)$-$^{12}_\Lambda\textrm{C}(3/2^-_2)$ as that in 
$^{11}\textrm{B}(3/2^-_2)$-$^{11}\textrm{C}(3/2^-_2)$. 
\label{fig:spe-yy2}}
\end{figure*}
%%%%%%%%%%%%%%%%%%%%%%%%%%%%%

%%%%%%%%%%%%%%%%%%%%%%%%%%%%%%
\begin{figure}[!h]
\begin{center}
\includegraphics[width=8.0cm]{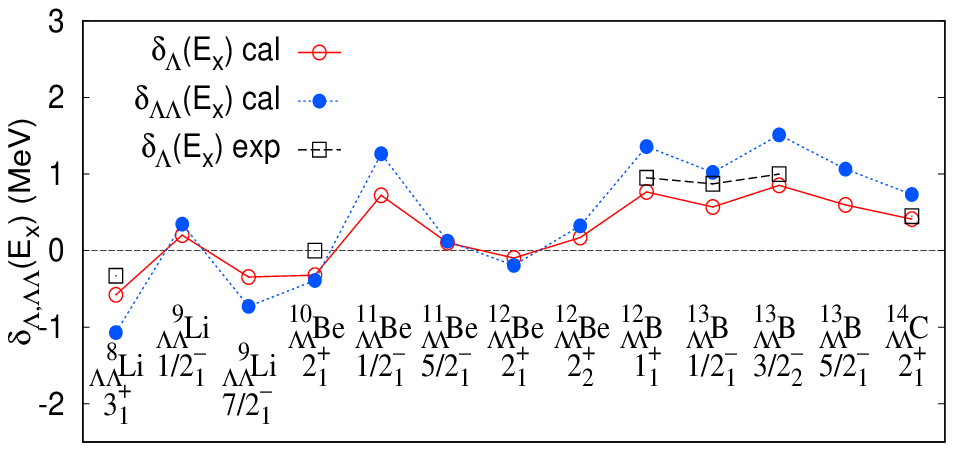} 	
\end{center}
%\vspace{0.5cm}
  \caption{(color online) 
Excitation energy shift $\delta_\Lambda(E_x)$ in $\LZ$ and $\delta_{\Lambda\Lambda}(E_x)$ in $\LLZ$ 
 calculated with ESC08a(Hyb).
The experimental values of $\delta_\Lambda(E_x)$  in $\LZ$   are also shown.
\label{fig:r-e-yy}}
\end{figure}
%%%%%%%%%%%%%%%%%%%%%%%%%%%%%

We discuss properties of excited states in $\LLZ$ 
such as the  $\La$-$\La$ binding and energy spectra. 
The calculated values of $\Delta B_\LLa$ and ${\cal V}^\textrm{bond}_\LLa$ 
obtained by ESC08a(Hyb), ESC08a(DI), and ESC08a(DD) 
are listed in Table \ref{tab:LL-ex}. 
Trends of system and interaction dependences 
for excited states in $\LLZ$  are similar
to those for the ground states. 

The calculated energy spectra in $^{A-2}Z$, $\LZ$, and $\LLZ$  are shown in Figs.~\ref{fig:spe-yy1} and 
\ref{fig:spe-yy2} compared with the observed 
energy levels in $^{A-2}Z$ and $\LZ$.
Figure \ref{fig:r-e-yy} shows the ESC08a(Hyb) result of 
excitation energy shift $\delta_{\Lambda\Lambda}(E_x)$ in $\LLZ$ 
compared with $\delta_\Lambda(E_x)$ in $\LZ$. Similarly to the case of$\LZ$, 
the excitation energy shift in  $\LLZ$ also correlates with 
the size difference between the ground and excited states.
The energy shift $\delta_\LLa(E_x)$ in  $\LLZ$ is roughly twice of $\delta_\La(E_x)$ in $\LZ$, except for $^{10}_\LLa \textrm{Be}$.
The result indicates that the core polarization effect is minor in the excitation energies
and each $\Lambda$ particle almost independently contributes to
the excitation energy shift through the $\Lambda N$ interactions.

\subsection{Comparison of $\Delta B_\LLa$ 
with experimental observations of double-$\Lambda$ hypernuclei}
%In the present calculation, in which there is no adjustable parameter in the $\La NG$ interactions,  
Experimental observations of double-$\Lambda$ hypernuclei are very limited and not enough to
discuss details of system dependence of the $\La$-$\La$ binding energies. 
We here compare our result with available data reported in 
experimental studies with nuclear emulsion \cite{Nakazawa:2010zzb}.

In the present calculation,  the
reproduction of $B_\Lambda$ is not precise enough to directly
compare the result with observed $B_\LLa$ values.
Alternatively, we compare the calculated $\Delta B_\LLa$ with 
the observed values. 
Note that, the calculated $B_\La(\LZ)$ values for $I\ne 0$ correspond to the spin-averaged values 
because the $\La$-spin dependent contributions are ignored in the present calculation, and hence, 
one should compare 
the calculated $\Delta B_\LLa(\LLZ)$ with the observed 
$\Delta B^\textrm{exp}_\LLa(\LLZ)$ reduced using the spin-averaged single-$\Lambda$ binding energy 
$\Delta \bar B^\textrm{exp}_\La(\LZ)$.

For $\LLBe$,  the DEMACHIYANAGI event has been observed and 
assigned to the excited state, $2^+_1$.
From the observed value  $B^\textrm{exp}_\LLa=11.90\pm 0.13$ MeV, $\Delta  B^\textrm{exp}_\LLa(\LLBe;2^+_1)$ is 
estimated to be $1.54\pm 0.15$ MeV 
using $E^\textrm{exp}_x(^{8}\textrm{Be};2^+_1)=3.04$ MeV for the resonance state
and the spin-averaged data $E^\textrm{exp}_x(^{9}_\La\textrm{Be};2^+_1)=3.05$ MeV.
Our result, $\Delta  B_\LLa(\LLBe;2^+_1)=1.49$ MeV,  of  ESC08a(Hyb) agrees to the 
experimental value.
There is no experimental data for the ground state, $\LLBe(0^+_1)$. 
 $\LLBe$ has been theoretically investigated 
with four-body calculations of the OCM $2\alpha+\LLa$ cluster model in Refs.~\cite{Hiyama:2002yj,Hiyama:2010zzd}. 
The latest calculation in \cite{Hiyama:2010zzd} successfully reproduces
$B^\textrm{exp}_\LLa(\LLBe;2^+_1)$ for the excited state, 
and predicts $\Delta B_\LLa(\LLBe;0^+_1)=1.32$ MeV for the ground state.
The present result of $\Delta B_\LLa(\LLBe;0^+_1)=1.74$ MeV obtained with ESC08a(Hyb)
is in reasonable agreement with their prediction.
However, the ESC08a(DI) (ESC08a(DD)) calculation gives 
much larger (smaller) $\Delta B_\LLa$ values than those of  ESC08a(Hyb) and seem to contradict the observed value and 
theoretical predictions in Ref.~\cite{Hiyama:2010zzd}.

For $^{11}_\LLa \textrm{Be}$ and $^{12}_\LLa \textrm{Be}$, a candidate event called the HIDA event has been observed.
For a  possibility of $^{11}_\LLa \textrm{Be}$, 
$\Delta B^\textrm{exp}_\LLa(^{11}_\LLa \textrm{Be})= 2.61\pm 1.34$  MeV was estimated using the
non-spin-averaged value $B^\textrm{exp}_\La(^{10}_\La\textrm{Be};J^\pi=1^-_1)=9.11\pm 0.22$ MeV.
The spin-doublet splitting between $J^\pi=1^-_1$ and $J^\pi=2^-_1$ in 
$^{10}_\La\textrm{Be}$ is considered to be as small as $<100$ keV in the observation and 
120 keV in the shell model estimation \cite{Gal:2011zr}.  If the shell-model value is used 
for an evaluation of the spin-doublet splitting, $\Delta B_\LLa(^{11}_\LLa \textrm{Be})=2.76\pm 1.34$ MeV
is reduced.
%If theoretical predictions of the spin-averaged value 
%$\bar B(^{10}_\La\textrm{Be})=8.97$ MeV of the five-body calculation in Ref.~\cite{Hiyama:2010zzd}
%and $\bar B(^{10}_\La\textrm{Be})=8.86\pm 0.11$ MeV of the shell model estimation in Ref.~\cite{Gal:2011zr} are used, 
%$\Delta B_\LLa(^{11}_\LLa \textrm{Be})= 2.89\pm 1.27$  MeV and 
%$\Delta B_\LLa(^{11}_\LLa \textrm{Be})= 3.11\pm 1.29$  MeV are reduced, respectively. 
The five-body cluster model calculation with the OCM in Ref.~\cite{Hiyama:2010zzd} predicted 
$\Delta B_\LLa(^{11}_\LLa \textrm{Be})=0.29$ MeV, which seems not consistent with the 
$^{11}_\LLa \textrm{Be}$ assignment  of the data.
Our result of ESC08a(Hyb) is $\Delta B_\LLa(^{11}_\LLa \textrm{Be})= 1.11$ MeV. 
An alternative interpretation of the HIDA event is a possibility of $^{12}_\LLa \textrm{Be}$ production.
For $^{12}_\LLa \textrm{Be}$, 
$\Delta B_\LLa$ can not be reduced because 
$B_\La(^{11}_\La \textrm{Be})$ is not known. In the shell model estimation with 
$\Lambda$-$\Sigma$ coupling and spin-dependent contributions, 
$\bar B_\La(^{11}_\La \textrm{Be})=10.02\pm 0.05$ MeV is predicted from 
$\bar B^\textrm{exp}_\La(^{11}_\La \textrm{B})$ \cite{Gal:2011zr}. Using the shell-model value, 
$\Delta B_\LLa(^{12}_\La \textrm{Be})=2.44 \pm 1.21$ 
MeV can be evaluated from $B^\textrm{exp}_\LLa(^{12}_\La \textrm{Be})=22.48\pm 1.21$ of the HIDA event. 
Our result of ESC08a(Hyb) is $\Delta B_\LLa(^{12}_\LLa \textrm{Be})= 0.72$ MeV. 
Because of the large uncertainty of the experimental data, we can not discuss agreement with data nor 
conclude which assignment is more likely.

For the $^{13}_\LLa \textrm{B}$ ground state, the experimental value $\Delta B^\textrm{exp}_\LLa=0.6\pm 0.8$ MeV
has been reported \cite{Nakazawa:2010zzb}. Our result $\Delta B_\LLa=0.60$ MeV of ESC08a(Hyb) is likely to be
consistent with the data.

\subsection{Size reduction in $\LZ$ and $\LLZ$}
In order to discuss the size reduction of core nuclei 
by $\La$ particles in $\LZ$ and $\LLZ$, we analyze ratios of nuclear sizes $R_N(\LZ)$ and $R_N(\LLZ)$  
to the original size $R_N(^{A-2}Z)$ defined as 
\begin{eqnarray}
S_\textrm{rmsr}(\LZ)&\equiv& \frac{R_N(\LZ)}{R_N(^{A-2}Z)}\\
S_\textrm{rmsr}(\LLZ)&\equiv& \frac{R_N(\LLZ)}{R_N(^{A-2}Z)}.
\end{eqnarray}
We also calculate $B(E2)$ values and the size reduction factor $S_{E2}$ from the ratios of 
$B(E2,\LZ)$ and $B(E2,\LLZ)$ in $\LZ$ and $\LLZ$ to the original value $B(E2,^{A-2}Z)$ defined as 
\begin{eqnarray}
S_{E2}(\LZ)&\equiv& \left [\frac{B(E2,\LZ)}{B(E2,^{A-2}Z)}\right]^{1/4},\\
S_{E2}(\LLZ)&\equiv& \left [\frac{B(E2,\LLZ)}{B(E2,^{A-2}Z)}\right]^{1/4}.
\end{eqnarray}
In the calculation of $B(E2,\LZ)$, we simply calculate 
the $E2$ transition strength for $I^\pi_i\to I^\pi_f$ in the core nuclear part 
while disregarding the spin coupling with $\La$s. 

The ESC08a(Hyb) result of nuclear sizes, $E2$ strengths,  and 
reduction factors are listed in Table \ref{tab:size}.
The significant size reduction by a $\Lambda$ particle occurs in $\LZ$ with the core nuclei, $^6\textrm{Li}$, $^7\textrm{Li}$ $^8\textrm{Be}$,
and $^9\textrm{Be}$,
because these nuclei have developed cluster structures, which
are fragile against the size reduction. In these clustered nuclei, 
further size reduction occurs by the second $\La$ particle in $\LLZ$.
The nuclear size reduction is $5\%\sim 25\%$ in $\LZ$ and $10\%\sim 30\%$ in $\LLZ$
for these light nuclei.
However,  in the case of heavier-mass core nuclei, $^{10}\textrm{Be}$, $^{10,11}\textrm{B}$, and $^{12}\textrm{C}$, 
the size reduction is as small as $2\%-4\%$ in $\LZ$ and $\LLZ$.

The size reduction by a $\La$ particle in $^7_\La\textrm{Li}$ has been investigated in experimental and theoretical studies.
The theoretical predictions of the OCM cluster model 
calculations are  $S_{E2}(^7_\La\textrm{Li})=0.83$ in Ref.~\cite{Motoba:1984ri} and 0.75 in Ref.~\cite{Hiyama:1999me}, which agree with 
the experimental value $S^\textrm{exp}_{E2}=0.81\pm 4$
reduced from the $B(E2)$ values in $^6\textrm{Li}$ and $^7_\La\textrm{Li}$  \cite{Tanida:2000zs}.
As discussed in the previous paper, we obtained 
$S_{E2}=0.74$ and $S_{E2}=0.86$ for $^7_\La\textrm{Li}$
in the ESC08a(DI) and ESC08a(DD) calculations, respectively. In the ESC08a(Hyb) calculation,  
an intermediate value $S_{E2}=0.82$ is obtained. The value is consistent with the experimental data and other calculations.

\begin{table*}[ht]
\caption{\label{tab:size} Properties of size reduction in $\LZ$ and $\LLZ$ calculated with ESC08a(Hyb):
 Nuclear size $R_N$, size change $\delta_\Lambda(R_N)$ and $\delta_\LLa(R_N)$ by $\Lambda$s, size reduction factors
$S_\textrm{rmsr}(\LZ)=R_N(\LZ)/R_N(^{A-2}Z)$ and 
$S_\textrm{rmsr}(\LLZ)=R_N(\LLZ)/R_N(^{A-2} Z)$ defined by sizes,  
$E2$ transition strength,
and the reduction factors $S_{E2}(\LZ)= [\frac{B(E2,\LZ)}{B(E2,^{A-2}Z)}]^{1/4}$ and 
$S_{E2}(\LLZ)= [\frac{B(E2,\LLZ)}{B(E2,^{A-2}Z)}]^{1/4}$ reduced from $B(E2)$. The units of sizes and $E2$ transition strengths are fm and $e^2$fm$^4$, respectively.
}
\begin{center}
\begin{tabular}{cccccccc}
\hline
core   ($^{A-2}Z$) & $J^\pi$ & $R_N(\LZ)$	& $\delta_\Lambda(R_N)$& $S_\textrm{rmsr}(\LZ)$  &  $R_N(\LLZ)$
& $\delta_{\Lambda\Lambda}(R_N)$ & $S_\textrm{rmsr}(\LLZ)$  	\\
$^6\textrm{Li}$	&	$1^+_1$	&	2.32 	&	$-0.24$ 	&	0.91 	&	2.22 	&$	-0.33 $	&	0.87 	\\
$^7\textrm{Li}$	&	$3/2^-_1$	&	2.34 	&	$-0.15$ 	&	0.94 	&	2.27 	&$	-0.23 $	&	0.91 	\\
$^8\textrm{Be}$	&	$0^+_1$	&	2.57 	&	$-0.80 $	&	0.76 	&	2.44 	&$	-0.93 $	&	0.72 	\\
$^9\textrm{Be}$	&	$3/2^-_1$	&	2.54 	&	-0.18 	&	0.94 	&	2.44 	&	-0.27 	&	0.90 	\\
%$^9\textrm{B}$	&	$3/2^-_1$	&	2.56 	&	$-0.20 $	&	0.93 	&	2.46 	&	$-0.30 	$&	0.89 	\\
$^{10}\textrm{Be}$	&	$0^+_1$	&	2.39 	&	$-0.06 $	&	0.97 	&	2.35 	&$	-0.10 $	&	0.96 	\\
$^{10}\textrm{B}$	&	$3^+_1$	&	2.34 	&	$-0.05 $	&	0.98 	&	2.31 	&	$-0.08 	$&	0.97 	\\
$^{11}\textrm{B}$	&	$3/2^-_1$	&	2.29 	&	$-0.04$ 	&	0.98 	&	2.26 	&	$-0.07 $	&	0.97 	\\
$^{12}\textrm{C}$	&	$0^+_1$	&	2.31 	&	$-0.04$ 	&	0.98 	&	2.28 	&	$-0.07 	$&	0.97 	\\
core ($^{A-2}Z$)	&	$I^\pi_i\to I^\pi_f$ &    $B(E2,^{A-2} Z)$	&	 $B(E2,^{A-2}Z)_\textrm{exp}$&	 $B(E2,\LZ)$		&	$S_{E2}(\LZ)$	&	 $B(E2,\LLZ)$		&	$S_{E2}(\LLZ)$\\
$^6\textrm{Li}$	&	$3^+_1\to 1^+_1$	&	11.3 	&	10.7(8)	&	5.0 	&	0.82 	&	3.7 	&	0.76 	\\
$^7\textrm{Li}$	&	$1/2^-_1\to 3/2^-_1$	&	19.6 	&	15.7(1.0)	&	12.4 	&	0.89 	&	9.5 	&	0.83 	\\
$^7\textrm{Li}$	&	$7/2^-_1\to 3/2^-_1$	&	11.0 	&	3.4	&	5.8 	&	0.85 	&	4.3 	&	0.79 	\\
$^8\textrm{Be}$	&	$2^+_1\to 0^+_1$	&		&		&	22.6 	&		&	15.3 	&		\\
$^9\textrm{Be}$	&	$5/2^-_1\to 3/2^-_1$	&	36.1 	&	24.4(1.8)	&	25.0 	&	0.91 	&	20.2 	&	0.87 	\\
%$^9\textrm{B}$	&	$5/2^-_1\to 3/2^-_1$	&	46.9 	&		&	30.2 	&	0.90 	&	23.2 	&	0.84 	\\
$^{10}\textrm{Be}$	&	$2^+_1\to 0^+_1$	&	11.7 	&	10.2(1.0)	&	9.6 	&	0.95 	&	9.2 	&	0.94 	\\
$^{10}\textrm{B}$	&	$1^+_1\to 3^+_1$	&	5.2 	&	4.15(2)	&	4.0 	&	0.94 	&	3.3 	&	0.89 	\\
$^{11}\textrm{B}$	&	$5/2^-_1\to 3/2^-_1$	&	9.5 	&	8.9(3.2)	&	8.2 	&	0.96 	&	7.3 	&	0.94 	\\
$^{12}\textrm{C}$	&	$2^+_1\to 0^+_1$	&	7.3 	&	7.6(4)	&	6.1 	&	0.96 	&	5.5 	&	0.93 	\\
\hline		
\end{tabular}
\end{center}
\end{table*}

\section{Summary}\label{sec:summary}

We investigated low-lying $0s$-orbit $\Lambda$ states 
in $p$-shell  double-$\Lambda$ hypernuclei
with microscopic cluster models for nuclear structure and a 
folding potential model for $\Lambda$ particles. 
Systematics of the energy spectra and $\LLa$ binding were discussed in relation with the 
nuclear core polarization. 
The reductions of the nuclear sizes and $E2$ transitions by $\La$ particles in $\LZ$ and $\LLZ$ were also discussed. 

We used the density-dependent 
effective $G$-matrix $\Lambda$-$N$ interactions with the ESC08a parametrization. 
As for the $k_f$  parameter of the density dependence in the $\Lambda NG$ interactions, 
we adopted three choices of ESC08a(DI), ESC08a(Hyb), and ESC08a(DD), which correspond to 
density-independent, mild density-dependent, and original density-dependent  interactions,
respectively. The ESC08a(Hyb) calculation consistently reproduces both
the $\Lambda$-$\Lambda$ binding energy in $^{10}_\LLa \textrm{Be}^*$ and the   
excitation energy shift in $\LZ$ with $A\ge 12$. 
However, ESC08a(DI) and ESC08a(DD) overestimates and underestimates
the observed data, respectively. 
 
We discussed the $\LLa$  binding in $p$-shell double-$\La$ nuclei focusing on system dependence of  
$\Delta B_\LLa$ (the $\La$-$\La$ binding energy) and ${\cal V}^\textrm{bond}_\LLa$ (the $\LLa$
bond energy). 
In the present result,  $\Delta B_\LLa$ shows significant system dependence, whereas 
${\cal V}^\textrm{bond}_\LLa$ is nearly independent from system (mass-number independent). 
The system-independent behavior of ${\cal V}^\textrm{bond}_\LLa$ is consistent 
with the results of OCM cluster model calculations 
in Refs.~\cite{Hiyama:2002yj,Hiyama:2010zzd} and supports their argument. 
The system dependence of $\Delta B_\LLa$ is dominantly described by the core polarization energy $\delta_\La(E_N)$.
In the light-mass nuclei, 
$^8_\LLa\textrm{Li}$, $^9_\LLa\textrm{Li}$, $^{10}_\LLa\textrm{Be}$, and $^{11}_\La\textrm{Be}$,  
significant deviation of $\Delta B_\LLa$ from the global systematics is found because of 
remarkable core polarization in the developed cluster structures. 
In $A\ge 12$ systems, the core polarization effect is minor, and 
$\Delta B_\LLa$ approaches ${\cal V}^\textrm{bond}_\LLa$ with increase of the mass number and 
shows only slight system dependence.
In order to extract clean information of the $\LLa$ binding 
from observation of binding energy of double-$\Lambda$ hypernuclei,
heavier-mass nuclei may be favored because they are more 
free from the core polarization effect rather than very light-mass ones.

In the present calculation, the system dependence of $\Delta B_\LLa$ comes dominantly 
from the core polarization. Other effects
might also contribute to $\Delta B_\LLa$, for example,  
higher partial waves of $\Lambda$s, 
spin dependence of the $\La$-$N$ interactions, 
$\Lambda$-$\Sigma$ coupling, which are ignored in the present calculation.
These effects may give additional contributions to $\Delta B_\LLa$.
For precise description of $\Delta B_\LLa$, improved 
calculations by considering these effects are requested. 

\begin{acknowledgments}
The author thanks to Dr.~Motoba and Dr.~Isaka for fruitful discussions.
This work was inspired by the  Karuizawa workshop (June 2017).
% organized by Kohno, 
%Nakamoto, Suzuki, and the author. 
The author would like to give a huge thanks to Dr.~Fujiwara for his continuous  
encouragement. 
%She also thanks to Dr.~Akaishi for many discussions 
%on hypernuclei and effective interactions during the time they were working in KEK. 
The computational calculations of this work were performed by using the
supercomputer in the Yukawa Institute for theoretical physics, Kyoto University. This work was supported by 
JSPS KAKENHI Grant Number 26400270.
\end{acknowledgments}

\end{document}